\documentclass[openacc]{rsproca_new}%%%%where rsproca is the template name

\usepackage{color}
\usepackage{gensymb}
\usepackage{amsmath}
\usepackage{dsfont}
\usepackage{xcolor}

\begin{document}

%%%% Article title to be placed here
\title{Post-buckling Dynamics of Spherical Shells}

\author{%%%% Author details
Marcel Mokbel$^{1}$,  Adel Djellouli$^{2,3}$,  Catherine Quilliet$^{2}$,  Sebastian Aland$^{1,4}$ and Gwennou Coupier$^{2}$}

%%%%%%%%% Insert author address here
\address{$^{1}$University of Applied Sciences (HTW) Dresden, Friedrich-List-Platz 1, 01069 Dresden, Germany\\
$^{2}$Universit\'e Grenoble Alpes, CNRS, LIPhy, F-38000 Grenoble, France\\
$^{3}$School of Engineering and Applied Sciences Cambridge, Harvard University, Cambridge, Massachusetts 02138, United States\\
$^{4}$Technische Universit\"at Bergakademie Freiberg, Akademiestrasse 6, 09599 Freiberg, Germany}

%%%% Subject entries to be placed here %%%%
\subject{Mechanics, Fluid Mechanics, Computational mechanics}

%%%% Keyword entries to be placed here %%%%
\keywords{Spherical shells, shallow shell theory, buckling, oscillators}

%%%% Insert corresponding author and its email address}
\corres{Gwennou Coupier\\
\email{gwennou.coupier@univ-grenoble-alpes.fr}}

%%%% Abstract text to be placed here %%%%%%%%%%%%
\begin{abstract}
We explore the intrinsic dynamics of spherical shells immersed in a fluid in the vicinity of their buckled state, through experiments and 3D axisymmetric simulations. The results are supported by a theoretical model that accurately describes the buckled shell as a two-variable-only oscillator. We quantify the effective "softening" of shells  above the buckling threshold, as observed in recent experiments on interactions between encapsulated microbubbles and acoustic waves. The main dissipation mechanism in the neighboring fluid is also evidenced.
\end{abstract}
%%%%%%%%%%%%%%%%%%%%%%%%%%%

%%%%%%%%%% Insert the texts which can accomdate on firstpage in the tag "fmtext" %%%%%

\begin{fmtext}
\section{Introduction}

Buckling of elastic structures has recently emerged as a powerful mechanism to trigger fast motion at small scale. This includes fast reorientation of microswimmers \cite{son13,huang2020}, thrust generation in fluids \cite{djellouli17,Wischnewski20} or, through solid friction \cite{Gorissen2020}, valves actuation for flow control \cite{gomez17_1,Rothemund2018} or fast actuation of optical lenses \cite{holmes07}. Smart design of materials in order to obtain the desired buckling behavior has now become an intense field of research \cite{yang15,holmes19,Stein-Montalvo19,janbaz19}.

\end{fmtext}

%%%%%%%%%%%%%%% End of first page %%%%%%%%%%%%%%%%%%%%%

\maketitle

From a modeling perspective, stable configurations of structures prone to buckling have been widely explored, as well as the stress or strain thresholds that have to be overcome to switch between states. The existence of (at least) two stable states separated by energy barriers allows for the design of robust devices that can maintain their state within a certain range of external perturbation without any external energy input. While the picture in terms of equilibrium states is now quite clear, at least for simple geometries (rods, half-spheres, closed spheres), full control of soft structures by external signals requires to know more about their dynamics. Recent papers have shed light on the complexity of the first-stage dynamics, close to the buckling threshold, where the response time of the material depends on classical dissipative mechanisms coupled to intrinsic slowing down observed in such a  critical phenomenon \cite{gomez17_2,Gomez19,Sieber2019}. The goal of the present paper is to explore the second-stage dynamics, around the buckled state, where the geometry is often more complex than that of the initial state.

Because of their ubiquitousness in nature but also of their simplicity in terms of fabrication and of modeling, spherical closed shells enclosing a compressible fluid are particularly in the spotlight \cite{carlson67,hutchinson67,berke68,landau86,quilliet08,quilliet10,knoche11,vliegenthart11,quilliet12,knoche14,djellouli17,hutchinson17,zhang17,Pezzulla2018,Coupier2019,li2021}. Existing studies are essentially focused on understanding the scenario of the buckling instability that occurs beyond a certain threshold of compression or deflation, and on characterizing the stability branches \cite{knoche11,quilliet12,knoche14,Coupier2019}. More recently, shells made of non isotropic material have also attracted some attention \cite{quemeneur12,Pitois2015,Munglani19}. In terms of dynamics, the reaction of shells to a steep increase of pressure have been recently studied \cite{Sieber2019,Stein-montalvo21} while an experimental study has highlighted the role of dissipation within the shell while reaching the stable buckled state \cite{Coupier2019}. 

Hollow microshells have been used for decades as ultrasound contrast agents (UCAs) and their resonance frequencies in the spherical configuration have been widely studied \cite{helfield19}. Noteworthy, even in such a simple configuration, the existing models lack to describe accurately all experimental observations \cite{chabouh21}. Depending on the applied acoustic field, UCAs may also buckle. In \cite{renaud15} the current state of a suspension of UCAs is controlled  by a low frequency acoustic field while the propagation velocity of  a high frequency acoustic signal is measured. The authors observe a decrease of this sound speed while the ambient pressure is increased above a certain threshold, in marked contrast with the standard results in a simple fluid. As in other preceding works  \cite{trivett06,zaitsev09}, this is interpreted as a "softening" of the shell due to its buckling. This interpretation is consolidated by the existence of a hysteretic loop as the ambient pressure is varied, which is also a signature of the buckling-unbuckling transitions. A very different study reaches the same conclusion: in  \cite{memoli18}, primary Bjerkness forces on hollow micrometric shells are measured ; a strong rise of this force above a given amplitude of the applied acoustic field is interpreted again as a signature of the sudden "softening" of the shell. This interpretation is supported by independent measurements of the buckling pressure by AFM.

In all above mentioned studies, the data are not quantitatively fitted by a model. Indeed, to our knowledge, the sole model accounting for shell response in the buckled state is that of Marmottant \textit{et al.} \cite{marmottant05},  which has been refined in \cite{sijl11}. These models assume that in the buckled configuration, the elastic response of the whole shell is simply that of the encapsulated gas while that due to shell material has disappeared, as if the shell was broken.

In the present work, we show that this rough approach is not valid. While our results confirm the effective softening due to buckling,  we highlight a more complex  interplay between gas response and shell material response, reaching the unexpected result that softening is even more pronounced than that obtained through neglecting shell material response.

\section{Statement of the problem}

\subsection{Description of relevant parameters}
 \label{subsec:param}

We consider spherical elastic shells made of an isotropic, incompressible elastic material of Young's modulus $E$ and initial thickness $d$. We have performed experiments on home-made shells of centimetric size (see Appendix \ref{sec:setup}), which allowed to check the consistency of our numerical simulations where the different relevant parameters could be varied on a wider range. In these simulations, we consider isolated zero-thickness elastic shells of initial radius $R_0$ whose elastic constants (compression and curvature modulus) are functions of $E$ and $d$ (see Appendix \ref{sec:method}). These shells are immersed in a Newtonian fluid in which the Navier-Stokes equation is solved. They are filled with a gas at pressure $P$ which is assumed to be instantaneously set by the shell volume according to an adiabatic process, a reasonable hypothesis considering the high velocities encountered in this problem: $P V^{\kappa}=P_0 V_0^{\kappa}$, where $V$ is the shell volume, $V_0$ its initial value, $P_0$ the initial pressure, and $\kappa$ is the polytropic coefficient. While the fabrication process of shells makes it difficult to obtain a measurable initial pressure other than the atmospheric one, the simulations have allowed to vary it so as to explore the relative contributions of gas compressibility and shell elasticity on the overall response. The thin shell limit that is considered here, though it may appear as a strong simplification, is indeed a relevant model even to describe thick shells, as evoked in \cite{Coupier2019} and confirmed in the following. The ambient pressure is initially equal to $P_0$ and is suddenly increased to a constant value ${P}_{\text{ext}}$, which is large enough to trigger buckling.

The large range of parameters we consider here  will allow us to establish post-buckling dynamics for a wide range of objects and scales, from the thin colloidal shells that are used, in particular, as UCAs \cite{marmottant05,marmottant11,errico15,segers20} or photoacoustics contrastagents \cite{Vilov17}, to macroscopic shells in elastomer which are among the favorite building blocks in soft robotics \cite{yang15,djellouli17,Gorissen2020}.

To describe this whole range, we consider the dimensionless problem obtained by considering the shell radius $R_0$ as the lengthscale (its midplane radius for a real shell), $E d/R_0$ as the pressure and elastic modulus scale, and $\rho_f$ as the density scale. The time scale is then $\sqrt{\rho_f R_0^3/(E d)}$. This scaling is that of the undamped period of an oscillating shell when the contribution of inner pressure is neglected.

The problem now depends on four parameters: the reduced thickness of the shell $\hat{d}=d/R_0$, the initial pressure   $\hat{P}_0=P_0/(E d/R_0)$ in the shell when it is in its stress-free spherical configuration, the applied pressure $\hat{P}_{\text{ext}}$ and the dimensionless viscosity of the fluid $\hat{\eta}=\eta_f/\sqrt{E d \rho_f R_0}$ that will characterize the damping in the system. This last number is the equivalent of an Ohnesorge number where surface tension has been replaced by the 2D elastic modulus $E d$. We detail in Table \ref{tab:orders} the typical values of the three parameters $\hat{d}$, $\hat{P}_0$ and $\hat{\eta}$ (characterizing the initial state) one may find when considering microshells and macroshells used within the current research context, as well as the range covered by our experiments and simulations.

Both in numerical simulations \cite{quilliet12,knoche11,hutchinson17} or in experiments \cite{Coupier2019}, it is now well established that at equilibrium, the pressure difference $P_{\text{ext}}-P$ quasi-plateaus to a constant $\Delta P_{pl}$ as a function of equilibrium volumes $V_e$. Therefore, varying the fourth parameter $\hat{P}_{\text{ext}}$ strictly amounts to varying the equilibrium pressure $\hat{P}_e$ inside the shell, which we already set here by varying the initial pressure. In our experiments and simulations, $\hat{P}_{\text{ext}}$ is typically chosen such that the pressure difference is right above the buckling threshold $\Delta \hat{P}_b = 4 \hat{d}/3$  \cite{hutchinson17,knoche11}.  Increasing too much the external pressure, or starting with very low pressure inside the shell, leads to a full collapse of the shell, with the two opposite poles being in contact. This pertains to a new kind of physics taking into account solid friction and requires additional development in the numerical method. We will avoid these extreme situations in the present work.

\begin{table*}[t]
\begin{center}
\tiny
\begin{tabular}{llllllllll}
\hline
Shell type & $R_0$  & $d$ & $E$  & $P_0$  & $\rho_f$  & $\eta_f$  & $\hat{d}$& $\hat{P}_0$ &$ \hat{\eta}$ \\
   & (mm) & & (MPa) & (MPa) & (Kg/m$^3$) &   (mPa$\cdot$s) & $(\times 10^{-3})$&  &$ (\times 10^{-3})$ \\
\hline
UCA& 0.001 - 0.005 & 2 - 5 nm & 1 - 100  & 0.1  & 1000  & 1  & 0.4 - 5 & 0.2 - 250 & 20 - 700 \\
Macroshell & 1 - 100  & 0.01 - 0.5$\, R_0$  & 0.01-1   & 0.1 - 10  & 1000  & 1 - 1000  &10 - 500 & 0.2 - $10^5$ & 0.004 - 3000 \\
\hline
Experiments & 6.75 - 67.5  & 0.08 - 0.3$\, R_0$ & 0.5 & 0.1 & 1260 & 1000 & 80 - 300 & 0.7 - 2.5 &  1.3 - 13 \\
Simulations &&&&&&&40 - 300 & 0.2 - 5 & 0.6 - 1300\\
\hline
\end{tabular}
\end{center}
\caption{\label{tab:orders} Some orders of magnitude for two typical objects of interest: ultrasound contrast agents (UCAs) \cite{helfield19,chabouh21} and polymeric macroscopic shells \cite{djellouli17,Gorissen2020}. Initial inner pressure in UCAs is often assumed to be close to atmospheric pressure while it may be varied on purpose in macroscopic shells. The ranges of  parameters covered by the present experiments and  simulations are also indicated. See text for the definition of the dimensionless parameters (hatted symbols).}
 \end{table*}

\begin{figure}[t]
	\centering 		
\includegraphics[width=\textwidth]{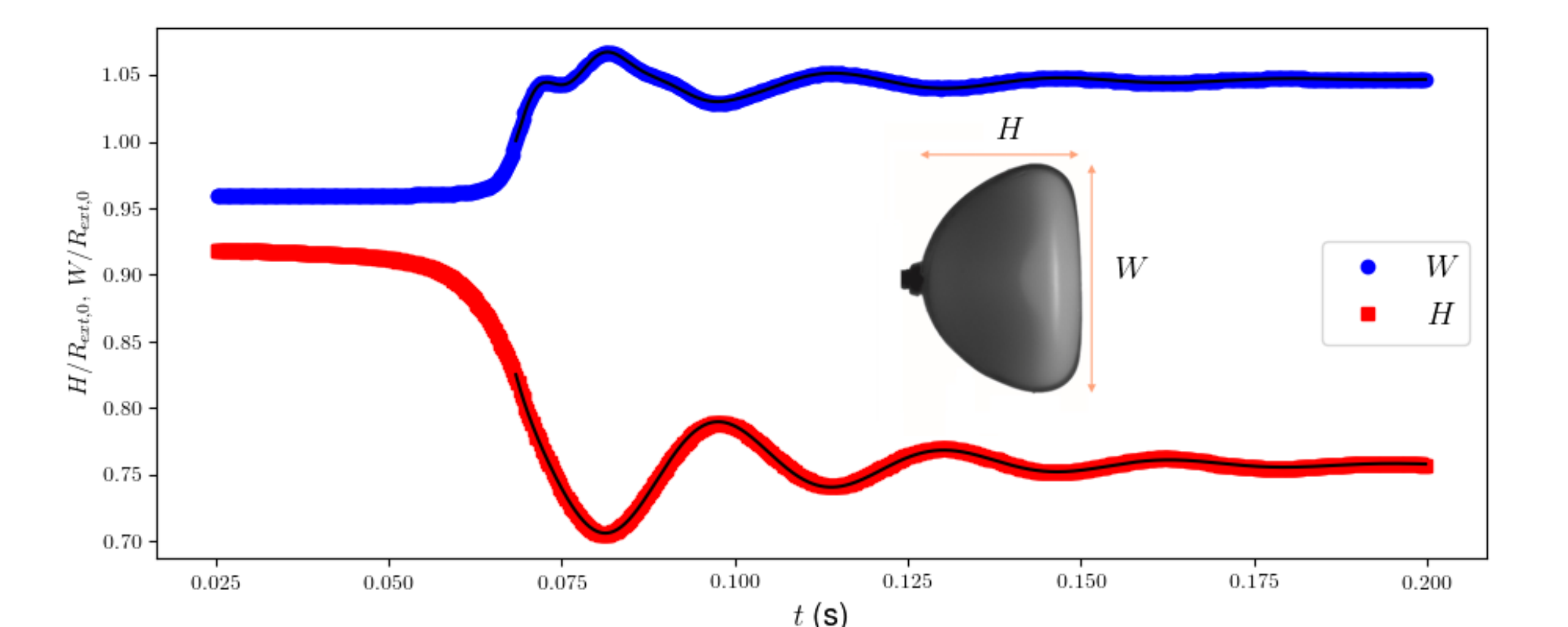}
	\caption[]{Oscillations of the width $W$ and height $H$ of the convex envelop of a shell with external radius $R_{ext,0}=25$ mm, thickness $d=5$ mm, Young modulus $E=0.5$ MPa,  initial internal pressure $P_0=1$ bar, external pressure ${P}_{\text{ext}}=1.77$ bar, immersed in glycerol (experiments in the reference configuration $\mathcal{R}$). Black lines correspond to the fit by the sum of two damped sinusoidal functions, which are obviously necessary to recover the full curve for the width $W$. The two frequencies are of the order 30 and 90 Hz, and are much lower than that typical of spherical configuration, which is  $f=\frac{1}{2 \pi R_0}\sqrt{\frac{1}{\rho_f} (3 \kappa P_0+ 4 E d/R_0)}\sim$ 210 Hz (see e.g. \cite{marmottant05}). Here $R_0$ is the midplane radius $R_{ext,0}-d/2$. Note that the sole gas contribution leads to a Minnaert frequency $\frac{1}{2 \pi R_0}\sqrt{\frac{3 \kappa P_0}{\rho_f}} \sim 140$ Hz which is also larger than the observed frequencies. These observations call for a finer modeling of the oscillation mechanisms around the buckled state.\label{fig:exptyp}}
\end{figure}

In the following, we will often discuss the effect of the four control parameters by varying them from a reference configuration (also chosen for Figs. \ref{fig:exptyp} and \ref{fig:simulationResults}), named $\mathcal{R}$ hereafter,  where $\hat{d}=0.22$, $\hat{P}_0=0.9$, $\hat{\eta}=0.004$ and $\hat{P}_{\text{ext}}=1.6$. This corresponds e.g. to  a macroscopic situation where $\eta= 1$ Pa.s (e.g. glycerol), $R_0=22.5$ mm, $d=5$ mm, $E=0.5$ MPa, $P_0=1$ bar and ${P}_{\text{ext}}=1.77$ bar (see Fig. \ref{fig:exptyp}) or to a microscopic configuration with the same pressures and elastic modulus  and   $\eta= 1$ mPa.s (e.g. water), $R_0=22.5$ $\mu$m,  and $d=5$ $\mu$m.

\subsection{A 2-frequency oscillator}
\label{subsec:2freq}

In the experiments as in the simulations, we consider the dynamics of a shell which is brought to the  onset of buckling by setting the external pressure  ${P}_{\text{ext}}$ to a value determined by a preliminary study. The external pressure is then kept fixed at this value and the dynamics of deformation of the shell is recorded. The sudden collapse of a shell at the buckling transition is followed by oscillations around the new equilibrium position. We show in Fig. \ref{fig:exptyp} the oscillations of two geometrical characteristics of a shell in configuration $\mathcal{R}$, corresponding to one of our experiments. A noticeable effect is the apparition of a second frequency in the oscillation pattern ; moreover, in agreement with the widely admitted softening of buckled shells, these two frequencies are much lower than that typical of the spherical state.

The apparition of a second frequency invalidates the previous models where the oscillations come from the sole contribution of gas compressibility. Note that we cannot expect more complex behavior, like the apparition of a second frequency, to emerge from a bubble yet having a non-spherical  shape: it has been shown recently \cite{Boughzala21}, in agreement with   \cite{Strasberg53}, that the generalized Minnaert model, where the radius of the spherical bubble is replaced by the effective radius extracted from the volume, is robust against geometry changes.

In the experiments, buoyancy issues have required  to restrict the displacement of the shell, which was attached to a fixed support through a suction pad  located at the pole opposite to the buckling pole.  In order to build up a general view of the oscillation mechanism devoid of any suspicion of strong influence of the boundary conditions, we turn to the numerical simulations which allow to consider free shells and a larger range of parameters. We shall come back in section \ref{sec:disc}\ref{sec:attached} to the impact of having a part of the shell that is fixed.

\section{Post-buckling oscillations of a free shell} \label{sec:simulresul}

 \begin{figure*}[t]
	\centering 		
	\includegraphics[width=0.96\textwidth]{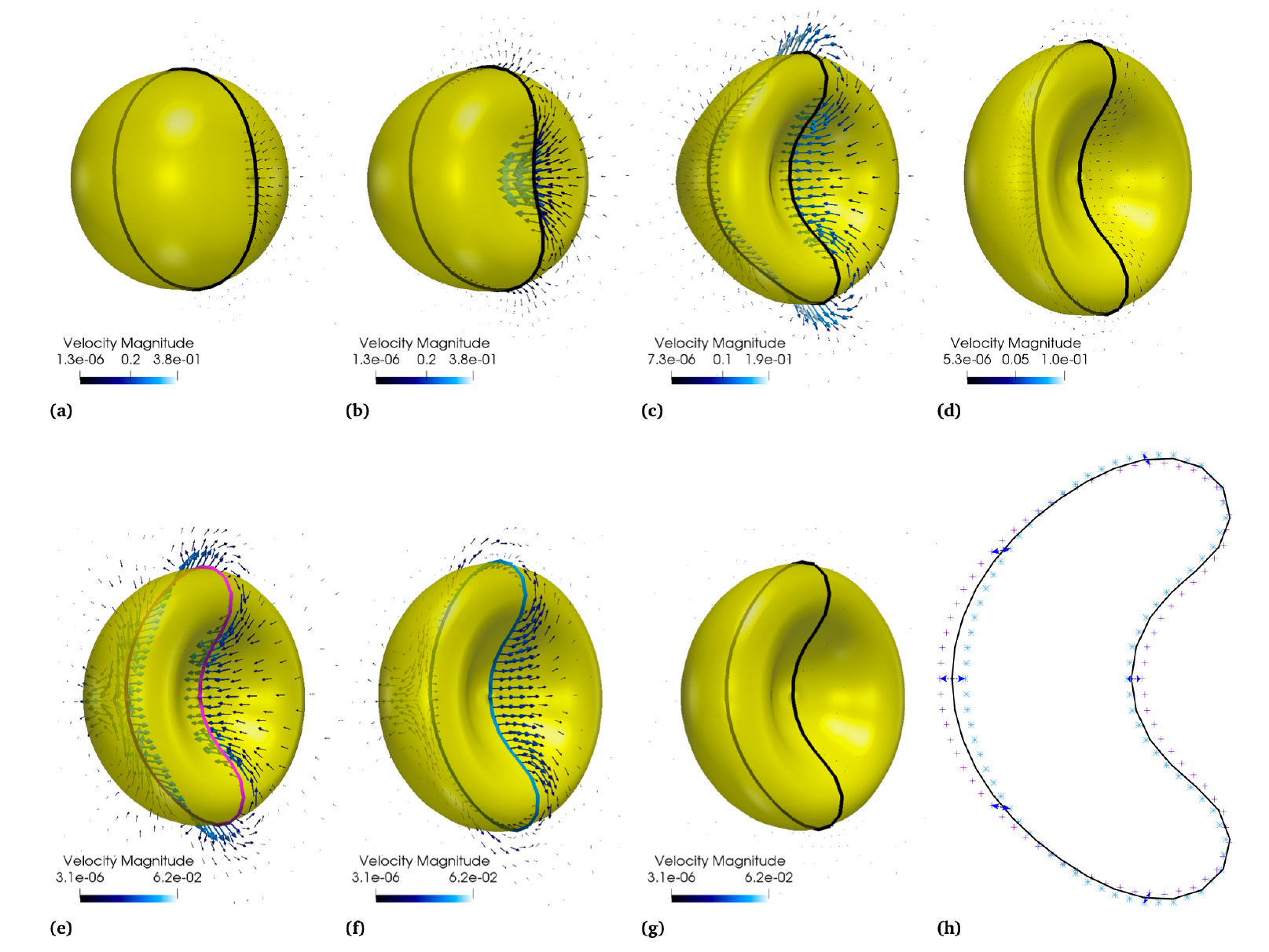}	
	\caption[]{Images of the buckling process from the simulations. The corresponding times are indicated in Fig. \ref{fig:oscill}(a). The arrows refer to the fluid velocity, scaled by the velocity magnitude. (a)-(d): buckling process, until the buckling spot reaches its deepest spot and the buckling oscillations around the equilibrium state set in. Two of these oscillating configurations are shown in (e)-(f). (g) is the final equilibrium state. (h) shows the respective contour of the shell, where the purple $+$ refer to the shape shown in (e) (illustrated by the purple shell contour), the $*$ refer to (f) (illustrated by the blue shell contour) and the black line to (g). The arrows indicate the movement of the surface grid points. The arrows in (e)-(g) are scaled 4 times larger than in (a)-(d) for visualization purposes. See also the corresponding movie in Supplementary Materials.\label{fig:simulationResults}}
\end{figure*}

 \begin{figure}[t]
	\centering 		
\includegraphics[width=\textwidth]{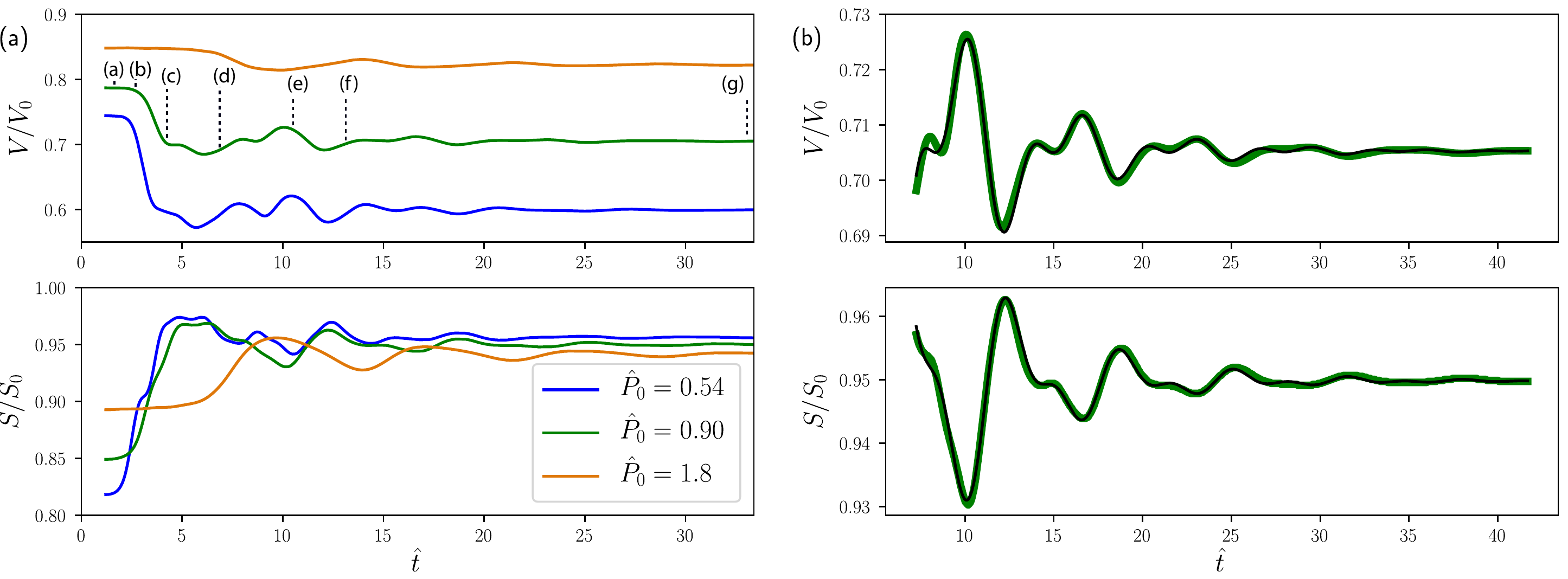}
	\caption[]{(a) Oscillations of the volume $V$ and surface $S$ of a shell in configuration $\mathcal{R}$ ($\hat{P}_0=0.9$) and two other initial pressures (simulations). Letters indicate the times at which the snapshots of Fig. \ref{fig:simulationResults} were taken. $V_0$ and $S_0$ are the initial values of the shell volume and surface, respectively. (b) Thin  black line : fit of the time evolution of $V$ and $S$ after the buckling stage for a shell in configuration $\mathcal{R}$, by the sum of two sinusoidal functions damped by the same decreasing exponential function. For both $V$ and $S$, the frequencies are similar.\label{fig:oscill}}
\end{figure}

We illustrate in Fig. \ref{fig:simulationResults} the buckling process as obtained from the simulations. From the many available data characterizing the shell shape, we focus on its volume $V$ and surface $S$ and plot in Fig. \ref{fig:oscill} their oscillations  towards their equilibrium values $V_e$ and $S_e$. The drop in volume is all the more pronounced as the initial pressure is low. By contrast, the shell surface  increases along the buckling process, and reaches a value close to the initial value of the sphere, which hardly depends on the initial pressure. This validates the vision of the buckling process as a mechanism that makes the shell switch from a spherically compressed state to a more favorable state where the in-plane  compression energy is released, the cost to pay being the curvature energy which is localized in the rim \cite{landau86}.

\begin{figure}[t]
	\centering 		
\includegraphics[width=0.8\textwidth]{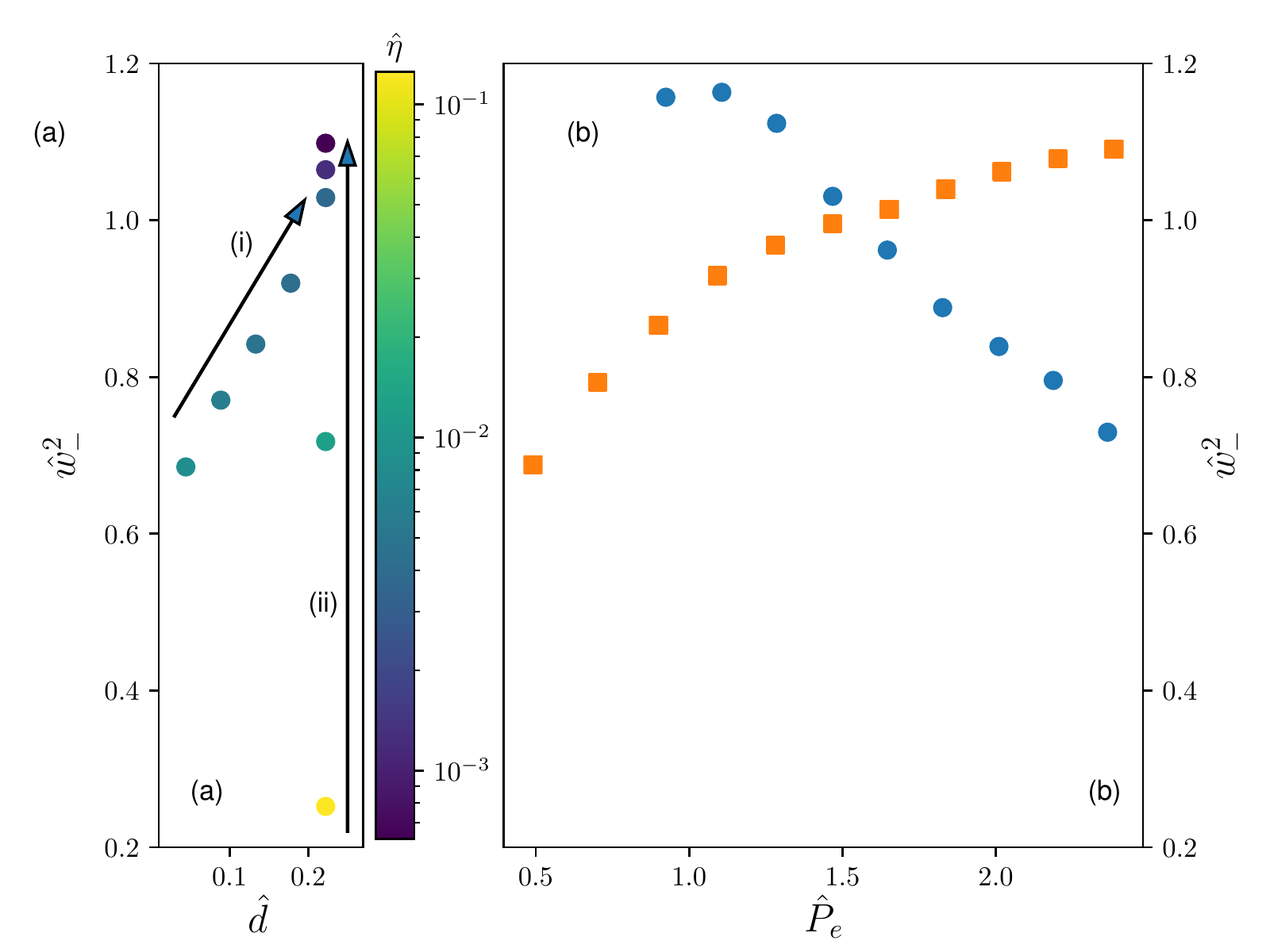}
	\caption[]{Main pulsation $\omega_-$ as a function of some control parameters. From the reference configuration $\mathcal{R}$ ($\hat{d}=0.22$, $\hat{P}_0=0.9$, $\hat{\eta}=0.004$), one parameter is varied at a time. (a) Following (i): effect of varying $d$: $\hat{d}$ therefore varies but also  $\hat{\eta}$ (see colorbar). Following (ii): effect of varying $R_0$ at fixed $\hat{d}$ (thus $\hat{\eta}$). (b) Effect of varying $\hat{P}_0$ at fixed $\hat{P}_{\text{ext}}-\hat{P}_0$ (blue disks) or $\hat{P}_0$ with $\hat{P}_{\text{ext}}$ set such that the final shape is preserved whatever the initial pressure (orange squares). \label{fig:resul}}
\end{figure}

Right after the buckling transition, the signals are, as in the experiments, well fitted by the sum of two sinusoidal functions damped by the same decreasing exponential, as shown in Fig. \ref{fig:oscill}(b). For each simulation, we then obtain three characteristic parameters of the shell dynamics: two frequencies $\omega_-$ and  $\omega_+$, with  $\omega_-<\omega_+$, and a single relaxation time $\tau$. We define $\omega_\pm$ as the intrinsic frequency, obtained from the measured frequency $\Omega_\pm$ through $\omega_\pm^2=\Omega_\pm^2+1/\tau^2$. These three parameters are similar for both $V$ and $S$, as well as for other characteristic parameters such as the width or the height of the shell (as discussed with  Fig. \ref{fig:detail-attached} further in the text). They all depend, a priori, on the 4 control parameters of the problem. In all the simulations we considered, the contribution of the highest frequency to the signal is between 50\% to 80\% smaller than that of the lowest frequency. We shall then focus first on this latter. In Fig. \ref{fig:resul} we show how this frequency $\omega_-$ varies with some control parameters, starting from the reference configuration $\mathcal{R}$. In Fig.  \ref{fig:resul}(a) the initial pressure $\hat{P}_0$ is kept fixed and the external pressure is set right above the threshold for buckling, and the dependency with space parameters is explored. Decreasing the thickness $d$ of a shell of given radius leads to the intuitive result that the frequency decreases, as a result of the decrease of the shell elasticity. For a given reduced thickness $\hat{d}$, Fig. \ref{fig:resul}(a) shows that the reduced frequency does not vary  significantly for small enough values of the reduced viscosity $\hat{\eta}$. When damping becomes more important, the frequency decreases, illustrating the threshold at which viscous stresses start to influence the dynamics of the shell beyond a mere linear damping.

Less intuitive is the variation of the frequency with the equilibrium pressure inside the shell, shown in Fig. \ref{fig:resul}(b). Starting from shell in configuration $\mathcal{R}$, we varied  the initial pressure $\hat{P}_0$ from 0.36 to 1.8. The resulting equilibrium pressure  $\hat{P}_e$ increases conjointly with this initial pressure. The result is that in the meantime, the frequency decreases. One can question this observation by considering, in a first approach, that the contribution of the gas to the frequency would scale like the Minnaert frequency $\omega_M$ of a free spherical bubble: $\omega_M^2 \propto P_e/V_e^{3/2}$, where we have simply replaced the usual radius of the original Minnaert expression by a generalized radius based on the volume, as in \cite{Strasberg53,Boughzala21}. Since the volume at equilibrium $V_e$ is an increasing function of internal pressure $P_e$, it is not clear a priori that the Minnaert frequency increases with pressure. In the spherical case, it can however be calculated that $V_e^{3/2}$ does not increase as quickly as $P_e$. In our case of buckled configuration, data from the simulations show also that such a Minnaert contribution increases with the equilibrium (or the initial) pressure. Therefore the origin of the decrease of $\omega_-$ with increasing internal pressure has to be found elsewhere.

In order to separate the direct effect of pressure on compressibility from its indirect effect through its influence on the shell volume, it is insightful to consider the following configuration: from the reference configuration $\mathcal{R}$ in its buckled equilibrium state, we have varied the inner and external pressure by the same amount $\delta\hat{P}$, which does not modify the shape of the shell. Then, using this state as the reference state for inner pressure calculation (i.e. $\hat{P} \hat{V}^\kappa=(\hat{P}_e+ \delta \hat{P}) \hat{V}_e^\kappa$), we apply a small perturbation to the shell and measure the induced oscillations, thus characterizing the pressure contribution to the elastic response, at fixed geometry.  As shown in Fig. \ref{fig:resul}(b), while this contribution is now increasing with the pressure, this increase of the frequency squared with the inner pressure is sublinear, in contrast with the spherical case.  Note that the procedure amounts to vary $\hat{P}_{\text{ext}}$ together with $\hat{P}_0$ such that the final shape is the same in all simulations. We checked, for a zero offset $ \delta \hat{P}$, that these oscillations with small amplitudes are similar to that obtained right after the shell has buckled, which initially implies larger deformations. This can be seen by the proximity between the two data points for $\hat{P}_e \simeq 1.5$ on Fig. \ref{fig:resul}(b).

In order to decipher these behaviors, we introduce a reduced model which is then used to fit the dataset obtained from the simulations.

\section{Model} \label{sec:model}

We describe the dynamics of the system through Lagrangian mechanics which is an 
 energy approach that allows to introduce the generalized coordinates that are the more relevant to the problem (see e.g. \cite{doinikov06} for such an approach in the case of spherical oscillations).  

In this frame, the dynamics is given by: 
\begin{equation} 
\frac{d}{dt} \frac{\partial L}{\partial \dot{q}_i}-\frac{\partial L}{\partial {q}_i}=-\frac{\partial D}{\partial \dot{q}_i}. \label{eq:Lag}
\end{equation} 
The Lagrangian function $L$ is defined as $L=T-U$ with $T$ and $U$ the kinetic and potential energies of the system, respectively. The Rayleigh dissipative function  $D$ allows to introduce non conservative forces in the formalism, providing they originate from viscous friction and that pressure drag is negligible. We shall see later that, due to the high buckling velocities, this assumption is fragile.
The $q_i$ and $\dot{q}_i$ are the generalized coordinates and velocities that characterize the state of the system.

The observation that the dynamics is well described by two frequencies calls for a modeling of the system as a pair of coupled oscillators. We then choose to describe the shell through two parameters, that naturally appear as relevant parameters in this problem: while the volume $V$ is intimately related to the gas behavior and is also the usual control or output parameter in shell buckling studies, the surface $S$ appears in the stretching contribution of the elastic energy. While in the spherical case both parameters are interdependent, in the buckled state that introduces at least one more degree of freedom they can vary independently, as depicted in Fig. \ref{fig:schememodel}.

    The modeling consists in writing how the elastic energy of the shell material and the potential energy of the inner gas depend on these two variables $V$ and $S$, as well as the kinetic energy and the dissipation due to viscous friction.

\subsection{Potential energy}

The potential energy can be written $U=U_g+U_{\text{ext}}+U_S$.  The internal energy $U_g$  of the gas reads, assuming an adiabatic behavior: \begin{equation}\label{eq:Ug} U_g=\frac{P (V) V}{\kappa-1}=\frac{P_0 V_0^{\kappa}}{(\kappa-1) V^{\kappa-1}}.\end{equation} The term $U_{\text{ext}}$ is the opposite of the work done by the external pressure on the shell, therefore \begin{equation} U_{\text{ext}}= P_{\text{ext}} V.  \label{eq:Upext}\end{equation} Finally, $U_S$ is the elastic energy in the shell. For a given value of external pressure, equilibrium conditions $\frac{\partial L}{\partial V}=0$ and $\frac{\partial L}{\partial S}=0$ lead to the following  set of equations that determine the equilibrium configuration $(V_e,S_e)$:
\begin{eqnarray}
P_{\text{ext}}-P(V_e)+\frac{\partial U_S}{\partial V}|_{(V_e,S_e)}&=&0, \\
\frac{\partial U_S}{\partial S}|_{(V_e,S_e)}&=&0.
\end{eqnarray}

As recalled before, at equilibrium the pressure difference $P_{\text{ext}}-P$ quasi-plateaus  to a constant $\Delta P_{pl}$ as a function of equilibrium volumes $V_e$. The plateau pressure $\Delta P_{pl}$  depends on shell elastic properties and thickness over radius ratio. While there is to our knowledge no available theoretical expression for it, an ad-hoc expression was proposed in \cite{Coupier2019} based on numerical simulations. 

Buckling is the mechanism by which the compressed spherical shell switches into an energetically more favorable configuration. Since bending energy is located in the rim, one can consider in a first approximation that at equilibrium, the surface in the buckled state is not compressed, as shown in Fig. \ref{fig:oscill}(a) where the surface area at equilibrium is only 5\% lower than the initial one. In the scheme of Fig. \ref{fig:schememodel} (a), it corresponds to an inversion of the right spherical cap towards the left pole. This implies that we can consider in first approximation that $S_e=4 \pi R_0
^2$, independently from $V_e$.  As a consequence of this and of the existence of the plateau pressure  $\Delta P_{pl}$, $\frac{\partial^2 U_S}{\partial V^2}|_{(V_e,S_e)}=0$. This assumption has been validated afterwards by letting the above partial derivative  be a free parameter in the fitting procedure of the data, that eventually lead to very small values compared to the other elastic constants of the problem. Around an equilibrium configuration, the elastic energy of the shell finally reads \begin{eqnarray} U_S&=&U_S(V_e,S_e)-\Delta P_{pl} (V-V_e) \nonumber \\
&+&A_{VS} (V-V_e)(S-S_e)+\frac{1}{2}A_{SS} (S-S_e)^2,  \label{eq:US}\end{eqnarray}

where $A_{VS}=\frac{\partial^2 U_S}{\partial V\partial S}|_{(V_e,S_e)}$ and $A_{SS}=\frac{\partial^2 U_S}{\partial S^2}|_{(V_e,S_e)}$. These parameters depend, a priori, on the equilibrium state.

As discussed above, we expect the variations of elastic energy with $S$ to be that of the stretching energy in the spherical case, since the curvature energy is localized in the rim. The elastic energy for a sphere of radius $R$ can be directly obtained from the general expressions  that are recalled in the Appendix \ref{sec:method} (Eqs. \ref{Eq:bend} and \ref{Eq:stretch}):
\begin{align}
 &   U_{S,\text{sph}}= U_{\rm stretch}+U_{\rm bend} \mbox{, with}  \nonumber \\ 
 & U_{\rm stretch}= 8 \pi K_A (1-\frac{R}{R0})^2 R^2 \mbox{  and   }  U_{\rm bend}=  8  \pi K_B (1-\frac{R}{R0})^2 . \label{eq:NRJsph}
\end{align}

We have introduced  the area bulk modulus $K_A = dE$  and the bending stiffness $ K_B = d^3E/9$. From $ d^2 U_{\rm stretch}/d S^2|_{S_e}=E d/S_e$ in the spherical case, we therefore postulate that $A_{SS}=E d/S_e$.

Similar reasoning is not possible for $A_{VS}$, as $V$ and $S$ are interdependent in the spherical case. We simply assume that the scaling $A_{VS}=E d /V_e \,k_{VS}$, where $k_{VS}$ is a dimensionless constant, will capture the dependency with space variables. Since there are two relevant space variables that are $R_0$ (or $S_e$) and $V_e$, other scalings could be possible. Complementary tests with the general scaling $A_{VS}=E d /(V_e^{\beta} R_0^{3(1-\beta)})  k_{VS}$  while proceeding with the fitting procedure of the data according to the model have indeed shown that $\beta\simeq 1$ is the best choice.

\subsection{Kinetic energy}

\begin{figure}[t]
	\centering 		
	\includegraphics[width=0.8\textwidth]{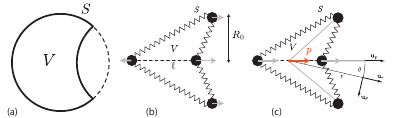}
	\caption[]{(a) Sketch of the parameters used to describe the shell dynamics. In addition to the volume $V$, we consider the surface $S$ of the shell. In the buckled configuration, both can vary independently. The simplified view with springs and masses presented in (b,c) allows to model the flow field resulting from the variations of $V$ and $S$. The Mises truss \cite{Huang72,Ario2004} is a convenient simplified model that has proven its efficiency to describe buckling  dynamics of shallow arch \cite{wiebe12} or elastic hemisphere (the "jumping popper" toy) \cite{Gomez19}, in particular when the system is subjected to oscillating external forces. In this paper, we do not go further in the analogy which will be developed in subsequent papers, but we simply make the remark that in the 3D shape obtained by axial rotation along the horizontal axis of the closed Mises truss depicted in (b) and (c), with the two lateral masses always located at fixed distance $2R_0$, $V$ variations can be made while keeping $S$ constant, as in (b), while in (c) surface variations are obtained while keeping $V$ constant. This is obtained by a similar horizontal motion of both poles, keeping their interdistance $\ell$ constant. This results in an hydrodynamic dipole  $\bf p$ at long range.  Its intensity is related to the volume changes on each side of the midsurface $S'=S/2$ indicated in grey. The assumption of fixed distance $2R_0$ between the two flanks is supported by the numerical simulations where we saw that most of the deformation occurs along the axis of symmetry (Fig. \ref{fig:simulationResults}(h)).\label{fig:schememodel}}
\end{figure}

We now turn to the determination of the kinetic energy $T$. The determination of the fluid flow around the oscillating shell is a complex problem, for which we only consider the long distance solution, obtained by adding the contributions of the constant surface and of the constant volume motions of the shell. 

Far from the shell (i.e. neglecting the shape details), shell volume variations lead to a flow induced by a point source, imposing a flow rate $q=\dot{V}$. By incompressiblity of the fluid, one obtains a first (radial) contribution ${\bf v}_1$ to the fluid velocity: \begin{equation} \label{eq:v1} {\bf v}_1=\frac{\dot{V}}{4 \pi r
^2} {\bf u}_r.
\end{equation}

Determining the flow due to surface variations is more complex. We wish to establish its dependency with the two dynamical variables $V$ and $S$. To that end, the geometry of the shell is simplified through an analogy with an axisymmetric spring-mass system, sketched in Figs. \ref{fig:schememodel}(b,c).

As sketched in Fig. \ref{fig:schememodel}(c), this simplification of the shell geometry leads to assimilate the deformation at constant volume to the synchronized motion of the two poles, thus considered as a pair of source/sink of flow rates $q'$ and $-q'$. This gives rise to an hydrodynamic dipole producing the flow field 
\begin{equation}
    {\bf v}_2= \frac{2 p \cos \theta }{4 \pi r^3} {\bf u}_r + \frac{ p \sin \theta }{4 \pi r^3} {\bf u}_{\theta},  \label{eq:v2}
\end{equation}
where $\theta$ is the angle between ${\bf u}_p$ that points in the pole to pole direction and the radial unit vector ${\bf u}_r$, and ${\bf u}_\theta$ is the associated unit vector (see Fig. \ref{fig:schememodel}(c)). The dipolar moment is $p=q' \ell$, where $\ell$ is the pole to pole distance. The dipolar approximation is well supported by the PIV observations made in \cite{djellouli17}, though it has been shown there that more complex patterns can take place due to shear waves near the flanks. The flow rate $q'$ is related to the deformation of the shell. As we have symmetrized the problem in a first approximation, by considering that the source and the sink have the same absolute intensity, we relate the flow rate $q'$ to the motion of the midsurface (of area $S'=S/2$) going through the midpoint between the two poles, depicted in grey on Fig. \ref{fig:schememodel}(c). As the poles move with velocity $v_p$, the rate of change of volume on each side of the mid-surface is  $q'/2=\pi R_0
^2 v_p/3$. A bit of geometry gives the relation between the pole velocity $v_p$ and the variation $\dot{S}'$ of the midsurface, eventually leading to \begin{equation} q'=-\frac{ R_0}{3} S (S^2-4\pi^2 R_0^4)^{-1/2}\dot{S}. \end{equation}

Finally, as $V=\pi R_0^2 \ell/3$, $p$ can be written under the form $p=g(V,S)\dot{S}$ where $g(V,S)=-\frac{V S}{ \pi R_0} (S^2-4\pi^2 R_0^4)^{-1/2}$.

Eventually, the kinetic energy is obtained by integrating $\frac{1}{2}\rho_f( {\bf v_1+v_2})^2$ on the whole fluid volume $\Omega$. By lack of knowledge on the effective shape, we integrate from $R_0$ to $\infty$ and compensate our successive approximations by introducing numerical prefactors $\alpha_1$  and $\alpha_2$ to the monopolar and dipolar terms, that we expect to be of order unity. They constitute the cost to pay in this  simplified - then tractable - model.

We eventually obtain 
\begin{equation}
%T&=&\frac{\rho_f}{2}\, 2 \pi\,  \int_0^{\pi}\int_{R_0}^\infty (\alpha_1 v_1^2 + \alpha_2 v_2^2) r^2 \sin \theta \,\mathrm{d}r \mathrm{d}\theta \nonumber \\
T=  \alpha_1\,\rho_f\, \frac{ \dot{V}^2}{8 \pi R_0} + \alpha_2\, \rho_f\,\frac{ g(V,S)^2 \, \dot{S}^2}{12 \pi  R_0^3} .
    \label{eq:T}
\end{equation}

\subsection{Dissipative function}

The dissipative function $D$ reads $D=\eta_f\int_{\Omega} (u_{ij})^2 dV$, where $u_{ij}$ is the symmetric strainrate tensor in the fluid. We implicitly assume here that dissipation originates from viscous friction and that pressure drag is negligible ; this is the framework in which the Rayleigh dissipative function has been introduced and validated so far. From the above expression for the fluid velocity, $D$ can be straightforwardly calculated and we find by integrating from $R_0$ to $\infty$ \begin{equation}\label{eq:D}
D=\beta_1  \, \frac{\eta_f}{2 \pi}\, \frac{\dot{V}^2}{R_0^3}+ \beta_2\,\frac{6 \eta_f}{5 \pi}\, \frac{g(V,S)^2\dot{S}^2}{R_0^5} , \end{equation}

where the $\beta_i$ have been introduced, as $\alpha_i$ for the kinetic energy, to account for the lack of precision in the integration domain.

\subsection{Dynamics around equilibrium}

We obtain the oscillation equations by considering only first order terms in $\Delta V= V-V_e$ and $\Delta S =S-S_e$ in the Lagrangian equation \ref{eq:Lag}, using Eqs. \ref{eq:Ug},  \ref{eq:Upext}, \ref{eq:US}, \ref{eq:T}, \ref{eq:D} and $g(V_e,S_e)^2=4 V_e^2/(3 \pi^2 R_0^2)$:

\begin{align}
\alpha_1 \frac{\rho_f}{4 \pi R_0} \Delta \ddot{V}=& -\frac{\kappa P_{e}(V_e)}{V_e} \Delta V - \frac{E d}{V_e} k_{VS} \Delta S - \beta_1 \eta_f\, \frac{1}{\pi R_0^3}\,  \Delta \dot{ V}\label{eq:eqmovV} \\
\alpha_2 \frac{2  \rho_f}{9 \pi^3 R_0^5} V_e^2 \Delta \ddot{S}=& - \frac{E d}{V_e} k_{VS} \Delta V - \frac{Ed}{4 \pi R_0^2}  \Delta S - \beta_2  \eta_f \,\frac{16 V_e^2 }{5 \pi^3 R_0^7}\, \Delta \dot{ S} \label{eq:eqmovS}
\end{align}

Together with the equilibrium conditions 
\begin{eqnarray} 
P_{\text{ext}}-P_e\,=\,\Delta P_{pl}&=& \frac{E}{\left(3/4\right)^{0.75}}\left(2.34\,10^{-6}+0.9\left(d/R_0\right)^{2.57}\right) , \nonumber \\ && \mbox{(according to \cite{Coupier2019})}   \label{eq:equiplateau}\\
P_0 V_0^{\kappa}&=&P_e V_e^{\kappa}, \label{eq:equivol}
\end{eqnarray} 
that set $P_e$ and $V_e$, these equations allow to determine the motion of the shell as a function of the initial pressure $P_0$ inside the shell, the external pressure $P_{\text{ext}}$, the initial shell radius $R_0$, its thickness $d$, Young's modulus $E$ and the fluid density and viscosity $\rho_f$ and $\eta_f$. The impact of the geometry  differs for each term considered in the equations, where $R_0$ and $V_e$ appear with different powers.

Switching to the dimensionless parameters, Eqs. \ref{eq:eqmovV} and \ref{eq:eqmovS} become:

\begin{align} 
\frac{\alpha_1 }{4 \pi } \Delta \ddot{\hat{V}}=& -\frac{\kappa \hat{P}_{e}(\hat{V}_e)}{\hat{V}_e} \Delta \hat{V} - \frac{k_{VS}}{\hat{V}_e} \Delta \hat{S} -  \hat{\eta}\, \frac{\beta_1}{\pi}\,  \Delta \dot{ \hat{V}}  \label{eq:eqmovadimV}\\
\frac{ 2 \alpha_2 }{9 \pi^3} \hat{V}_e^2 \Delta \ddot{\hat{S}}=& - \frac{k_{VS}}{\hat{V}_e}  \Delta \hat{V} - \frac{1}{4 \pi }   \Delta \hat{S} - \hat{\eta} \,\frac{16 \beta_2 \hat{V}_e^2 }{5 \pi^3}\, \Delta \dot{ \hat{S}}, \label{eq:eqmovadimS}
\end{align}

%Disregarding in Eq. \ref{eq:eqmovadimV} the coupling term in $\Delta \hat{S}$, one obtains a classical equation for a damped oscillator, whose quality factor $Q$ is close to $\hat{\eta}^{-1}$, assuming that $\alpha_i$ and $\beta_i$ are close to unity, as well as $\hat{P}_e$. 

%This scaling for the quality factor has important consequences regarding applications at the microscale: we rewrite first $Q$ as $Q=R_0\sqrt{\rho_f\Delta P_b \frac{R_0}{d}}/{\eta_f}$, where $\Delta P_b \sim E \left(\frac{d}{R_0}\right)^2$ is the buckling pressure. One can see that in water, for $R_0=1\,\mu$m, $d=10$ nm, and a buckling pressure of order 1 bar, the quality factor is of order 100. This means that the fluid does not influence the shell dynamics, which would simplify the design and optimization of such a system, as well as its modeling. 

In most cases of interest (see Table \ref{tab:orders}), $\hat{\eta}$ is small. This means that the fluid does not influence the shell dynamics, which would simplify the design and optimization of such a system, as well as its modeling: we can  consider $\hat{\eta}$ as a small parameter and look for solutions of Eqs. \ref{eq:eqmovadimV} and \ref{eq:eqmovadimS}  under the form $A e^{(i \hat{\omega}_0 -\hat{\delta}) t}$, where $\hat{\delta}$ is a small parameter, instead of a general form $A e^{i \hat{\omega} t}$ which would result in a complex quartic equation for $\hat{\omega}$. There are such non-trivial solutions of the equations of motions if and only if the determinant of the associated matrix is zero:
\begin{align}
\Big( \frac{\alpha_1 }{4 \pi }(i \hat{\omega}_0 -\hat{\delta})^2 +\frac{\kappa \hat{P}_{e}(\hat{V}_e)}{\hat{V}_e} +   \hat{\eta}\, \frac{\beta_1}{\pi}\, (i \hat{\omega}_0 -\hat{\delta}) \Big) \times &   \nonumber \\ 
\Big(
\frac{ 2 \alpha_2 }{9 \pi^3} \hat{V}_e^2 (i \hat{\omega}_0 -\hat{\delta})^2  + \frac{1}{4 \pi } + \hat{\eta} \,\frac{16 \beta_2 \hat{V}_e^2 }{5 \pi^3}\, (i \hat{\omega}_0 -\hat{\delta}) \Big)- \Big(\frac{k_{VS}}{\hat{V}_e}\Big)^2&=0.\label{eq:eqw}
\end{align}

Taking the leading order in $\hat{\delta}$ and $\hat{\eta}$ (which is the 0th order) of the real part of the above equation we obtain two eigenpulsations

  \begin{equation} \label{eq:defw}
   \hat{\omega}_{\pm}^2= \frac{ 9 \pi^2 \alpha_1  +   32 \pi \alpha_2  \kappa  \, \hat{V}_e  \,   \hat{P}_e\pm  \sqrt{(9 \pi^2 \alpha_1  -   32 \pi \alpha_2 \kappa  \, \hat{V}_e  \,   \hat{P}_e)^2+2\, \alpha_1\alpha_2 (48 \pi^2 \, k_{VS})^2 }}{16 \alpha_1\alpha_2 \hat{V}_e^2}.
\end{equation}

This constitutes the central result of this study. The leading order of the imaginary part of Eq. \ref{eq:eqw} gives the two relaxation times $\hat{\tau}_{\pm}=\hat{\delta}_{\pm}^{-1}$ associated with the two above pulsations. A priori, these two times can be different. Nevertheless, we directly make use of the observation that all obtained oscillation signals in the simulations are very well fitted by the sum of two sinusoidal functions damped by the same exponential function. Consequently, there is only one damping time. It can be shown that $\hat{\tau}_+-\hat{\tau}_-$ is proportional to $5 \alpha_2\beta_1-18 \alpha_1 \beta_2$, which is therefore taken to be 0. Injecting the obtained value of $\beta_2$ in the expression for $\hat{\tau}\equiv\hat{\tau}_{\pm}$, we obtain

\begin{equation} \hat{\tau}=2  \frac{\alpha_1}{\beta_1} \hat{\eta}^{-1} \qquad \mbox{that is, } \label{eq:taureal}\tau=2\frac{\alpha_1}{\beta_1} \frac{\rho_f}{\eta_f}R_0^2.\end{equation}

\begin{figure}[t]
	\centering 		
	\includegraphics[width=0.8\textwidth]{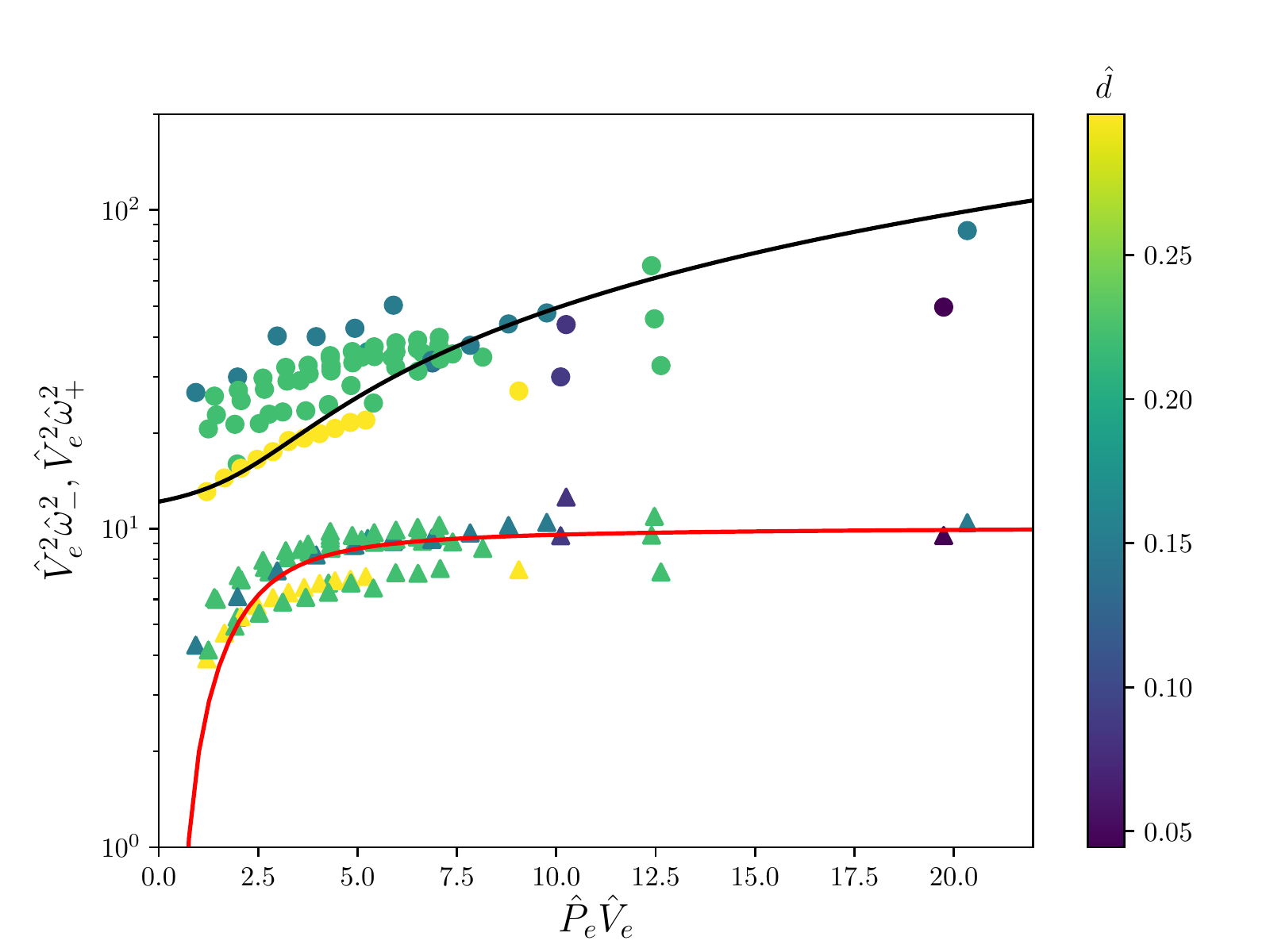}
	\caption[]{Dots : The two oscillations frequencies $\omega_-$ (triangles) and $\omega_+$ (disks) as determined from the fit of the oscillations from numerical simulations. A joint fit of the two set of data by Eq. \ref{eq:defw} leads to the full line curves.  \label{fig:w}}
\end{figure}

\section{Discussion}\label{sec:disc}

\subsection{Oscillation frequencies}
\label{subsec:agree}

We discuss the agreement between the model and the simulations. According to these, for the maximal value explored $\hat{\eta}=1.3$, the dynamics is fully damped. Below $\hat{\eta}=0.13$, few oscillations are seen and we are only able to determine the frequency $\hat{\omega}_-$ that contributes the most to the signal. Below $\hat{\eta}=0.013$ both frequencies can be accurately determined. This range corresponds also to the small damping limit that lead to Eq. \ref{eq:defw}. This equation suggests that $\hat{V}_e^2 \hat{\omega}_\pm^2$ plotted as a function of $\hat{V}_e \hat{P}_e$ follows a mastercurve. In Fig. \ref{fig:w}, we follow this idea and find that indeed the data collapse reasonably  on a single curve considering the large range of control parameters considered.  The set of data  for $\hat{V}_e^2 \hat{\omega}_-^2$ and $\hat{V}_e^2 \hat{\omega}_+^2$ are well fitted by the proposed equation (Eq. \ref{eq:defw}), the free parameters being $\alpha_1$, $\alpha_2$,  and $k_{VS}$.

%Note that the solution for the fit is not unique: any positive value of $\alpha_2$ can give the correct expression with an adequate adjustment of the three other free parameters $\alpha_1$, $k_{SS}$ and $k_{VS}$:  knowing the oscillation frequency of a mass-spring system is not sufficient to determine the unknown mass and spring stiffness. Here, we are in an intermediate situation where  the elastic contribution of the gas is fully known, and our data allow us to determine three constants. We choose to fix $\alpha_2=\alpha_1$.

Fitting the two frequencies together leads to $\alpha_1=3.6 \pm 3.5\%$, $\alpha_2=1.1 \pm 4.1\%$ , and $k_{VS}=0.23 \pm 6.6\%$. These are acceptable ranges: $\alpha_i$  larger than 1 was expected since it is intended to be a correction of the fact we haven not included the fluid inside the sphere of radius $R_0$ in the calculation of the kinetic energy.

%As discussed above, $k_{SS}$ was expected to be close to 1 as in the spherical case. Finally, finding $k_{VS}$ between $k_{SS}$  and  0, which would be the elastic energy cost in the shell material for volume-only variations, is a coherent result.

Interestingly, $\omega_-^2$ vanishes for a finite pressure, which sets a stability limit of the buckled state, which is given by \begin{equation}\label{eq:limstab}
\hat{V}_e \hat{P}_e = 4\pi \frac{k_{VS}^2}{\kappa }.
\end{equation}

 This provides the minimal pressure needed to maintain the buckled shape. 

The physical meaning of the two frequencies is particularly clear in the  large $\hat{P}_e$ limit.  From Eq. \ref{eq:eqmovadimV}, we see that  $\omega_+$  and $\omega_-$ can then be interpreted as the contributions of the  (now decoupled) volume and surface oscillations, respectively. From Eq. \ref{eq:defw}, we obtain
\begin{equation} \hat{\omega}_+^2  \underset{\hat{P}_e\to \infty}{\sim} \frac{4 \pi \kappa \hat{P}_e}{\alpha_1 \hat{V}_e}. \label{eq:MinBuck}
\end{equation}

Though this expression ressembles a Minnaert contribution, as in the model of Marmottant  \textit{et al.} \cite{marmottant05}, it indeed differs from it by a factor $R_0/(\alpha_1 R_e)$ and will be smaller in most cases, as illustrated in the next section.

The smallest pulsation $\omega_-$ becomes, in the large $\hat{P}_e$ limit: \begin{equation}
    \hat{\omega}_-^2  \underset{\hat{P}_e\to \infty}{\sim} \frac{9 \pi^2}{8 \alpha_2 \hat{V}_e^2}. \label{eq:omega-lim}
\end{equation}

Contrary to the spherical case, where the remaining contribution at high pressure is that of the gas pressure, we end up here with the sole contribution of the shell material compressibility. The high pressure makes the whole shell incompressible, but the buckled geometry makes it possible for the surface to evolve independently, with its own associated stiffness $A_{SS}$. The equivalent picture is that of two springs in series, one of them becoming infinitely stiff: the dynamics is then given by the smooth one.  This contrasts with the case of of the spherical geometry, where the springs  are in parallel (see Eq. \ref{eq:wsphere}). 

The above considerations help explaining the behavior shown in Fig. \ref{fig:resul}(b): as $\hat{P}_0$ increases, $\hat{V}_e$ increases and Eq. \ref{eq:omega-lim} shows that the pulsation is eventually a decreasing function of the initial - or of the equilibrium - pressure. For fixed volume $\hat{V}_e$ and varying inner pressure, Eq. \ref{eq:omega-lim}  tells us that the frequency converges towards a finite limit, thus explaining why it does not increase linearly with the pressure (as in the spherical case), as shown in Fig. \ref{fig:resul}(b).

\begin{figure}[t]
	\centering 		
	\includegraphics[width=\textwidth]{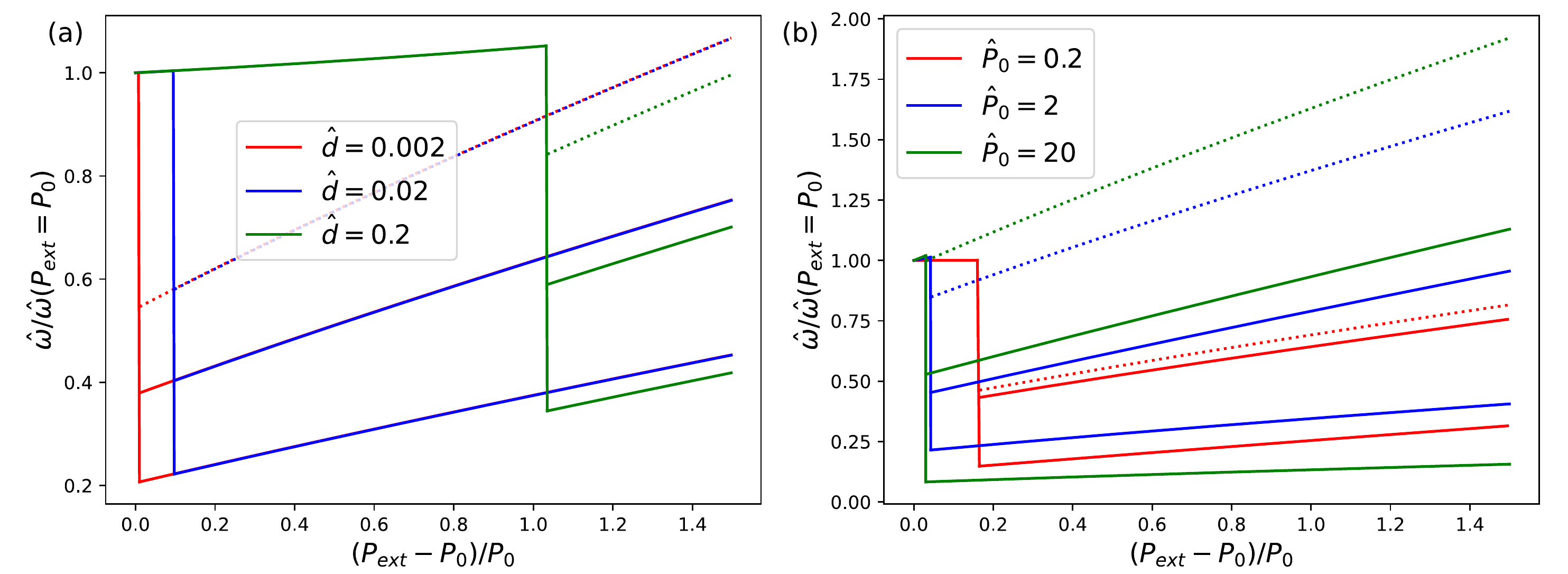}
	\caption[]{Oscillations frequencies around equilibrium position (normalized by that of the stress-free sphere) for several pairs of parameters $(\hat{d},\hat{P}_0)$, as a function of applied external pressure $\hat{P}_{ext}$. (a) $\hat{P}_0=0.4$ fixed and varying $\hat{d}$. (b) $\hat{d}=0.02$ fixed and varying $\hat{P}_0$. The low bounds correspond typically to commercial lipidic UCAs while the high bounds are relevant for soft polymeric shells which are often thicker but made of a softer elastic material. Full lines indicate the single frequency in the spherical case (see Eq. \ref{eq:wsphere} in the Appendix), and the two frequencies in the buckled case. Dotted lines correspond to the predictions in \cite{marmottant05} where it is assumed that   the pulsation is the Minnaert pulsation for the shell of radius $R_0$ at ambient pressure: $\omega=\frac{1}{ R_0}\sqrt{\frac{3 \kappa P_{ext}}{\rho_f}}$. \label{fig:soft}}
\end{figure}
\subsection{"Softening" of buckled shells}
\label{subsec:soft}
Our results demonstrate that the oscillation frequencies of a shell drop after it has buckled. This so-called "softening" is illustrated in  Fig. \ref{fig:soft} where we compare the obtained frequencies with that of the spherical case before buckling and that proposed in Ref. \cite{marmottant05} for the buckled case. To do so, we consider increasing values of the external pressure $P_{ext}$ starting at $P_0$ such that the shell is first in its spherical configuration until the pressure difference reaches the buckling pressure. In the buckled state, we consider either the Marmottant et al. model \cite{marmottant05} or ours. The equilibrium configurations are calculated for each value of $P_{ext}$, with the internal pressure given by the adiabatic condition (Eq. \ref{eq:equivol}) and the resulting pressure difference being equilibrated by the elastic restoring force, which is given by Eq. \ref{eq:equiplateau} in the buckled case and by $- dU_{S,\text{sph}} / d V$ in the spherical case, where $U_{S,\text{sph}}$ is given by Eq. \ref{eq:NRJsph}.

We show in particular that the "softening" is characterized by a drop of the oscillation frequencies by factors 5 (for $\omega_-$) and more than 2  (for $\omega_+$)  for parameters that are typical of the usual commercial UCAs, and is much more pronounced than that predicted in previous models, over a large range of parameters. \\

\begin{figure}[t]
	\centering 		
	\includegraphics[width=0.7\textwidth]{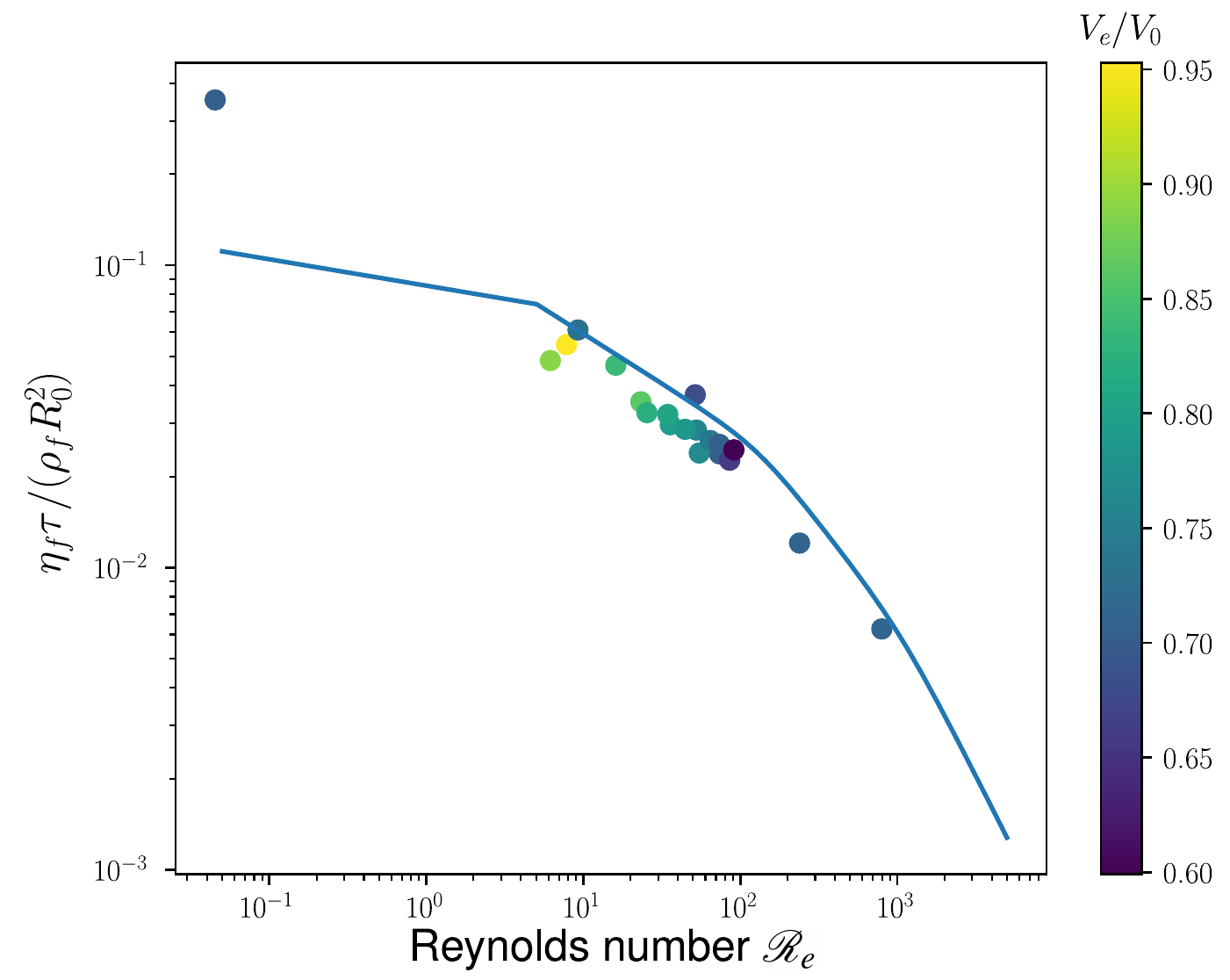}
	\caption[]{Theoretical oscillation relaxation time as a function of local Reynolds number (simulations). For a spherical bubble of radius $R_0$  settling in a fluid, the (velocity dependent) typical time needed to reach stationary velocity is  $\eta_f \tau/(\rho_f R_0^2)=8/(3 C_x \mathcal{R}_e)$, where $C_x$ is the  drag coefficient \cite{almedeij08} (that depends on $\mathcal{R}_e$) and $\mathcal{R}_e$ the Reynolds number. The full line indicates this quantity, which matches perfectly the measured damping time for the oscillations without any adjusted parameter. \label{fig:tau}}
\end{figure}

\subsection{Damping}
\label{subsec:damp}
The scaling of the damping time $\tau=2\frac{\alpha_1}{\beta_1} \frac{\rho_f}{\eta_f}R_0^2$ predicted by Eq. \ref{eq:taureal} is the only one that can be built when considering only viscous effects in a fluid of density $\rho_f$, viscosity $\eta_f$ in the vicinity of an object of size $R_0$. Indeed, the assumption of viscous friction only was needed to calculate a dissipative function.
It turns out that, due to the high Reynolds numbers that characterize the flow, this scaling  does not hold in most of the considered range of parameters. We consider the Reynolds number based on the maximal flow velocity $v_{\text {pole}}$, that is located near the buckling pole, which is then defined as $\mathcal{R}_e=v_{\text {pole}} R_0 \rho_f/\eta_f$. As shown in Fig. \ref{fig:tau}, the dimensionless relaxation time $\eta_f \tau/ (\rho_f R_0^2)$ indeed depends on the Reynolds number, while it would be constant if the dissipation was only of viscous origin (see Eq. \ref{eq:taureal}). The dependency with the Reynolds number is indeed pretty close (with no adjusted parameter) to that of the relaxation time towards constant velocity for a settling shell of radius $R_0$, for which the drag coefficient $C_x$ is known \cite{almedeij08}. This leads to the following general expression for the relaxation time: \begin{equation} \tau=\frac{8}{3}\frac{\rho_f R_0^2}{\eta_f C_x(\mathcal{R}_e) \mathcal{R}_e}.\end{equation}

The scaling of Eq. \ref{eq:taureal}, that was valid only in the low Reynolds limit, is preserved, but a dependency of the prefactor with the Reynolds number must be introduced, in order to account for the hydrodynamics regime. In particular, in the heart of our space parameter of interest, the dissipation regime is thus that of the Allen regime \cite{Goossens19}, where a combination of viscous and pressure effects leads to a drag coefficient that roughly scales like  $\mathcal{R}_e^{-0.6}$, in the $1 <\mathcal{R}_e<10^3$ range.

As for the purely viscous case, considering the data for $\hat{\eta}=0.13$, where  $\mathcal{R}_e<0.1$, Eq. \ref{eq:taureal} gives $\alpha_1/\beta_1=0.17$, quite close to the case of spherical oscillations where it would be equal to 0.125 according to  \cite{doinikov06}.

We eventually make the remark that the model predicts that, for not too high $\hat{\eta}$ - which will be the case in most situations - the frequencies scale with $1/R_0$, for a given reduced thickness $d/R_0$. This implies that the typical maximal velocity  at the pole $v_{\text {pole}}$, that scales like $\omega R_0$, is scale independent. In particular, on a large range of inner pressures, we can use the large pressure limit of Eq. \ref{eq:omega-lim} as an estimate for the leading frequency $\omega_-$, leading to 

\begin{equation}
v_{\text {pole}} \simeq \sqrt{\frac{9 \pi^2 E \frac{d}{R_0}}{8 \alpha_2 \rho_f \hat{V}_e^2}}.
\end{equation}

With typical values $ E \frac{d}{R_0}$ =  1 bar, $\rho_f=1000$ Kg/m$^3$ and $\hat{V}_e=1/2\times 4\pi/3 $, we find $v_p\sim 15$ m$\cdot$s$^{-1}$. This implies that, even at small scales, the Reynolds number will be quite high, in the range $1-10^3$ discussed above, that was obtained in the simulations.

%\begin{figure}[!h]
%	\centering 		
%	\includegraphics[width=0.6\textwidth]{Fig8}
%	\caption[]{Small frequency $\omega_-$ for 6 attached shells considered in experiments (disks) and corresponding simulations (squares). See Methods section for the experimental parameters. The crosses show the values found in the simulations for the 4 shells of intermediate radius, when they are free in the fluid. \label{Fig:compsimexp}}
%\end{figure}

\subsection{Fixed shells: the effect of boundary conditions}\label{sec:attached}

In practical applications, a shell may be linked to other components. For our experiments on six different shells (see Appendix \ref{sec:setup}), buoyancy issues have required  to restrict the displacement of the shell. For the three shells of  external radii 7.5 mm, 25 mm and 75 mm, but identical $\hat{d}$ we checked the scalings $\omega \propto 1/R$. We have also simulated the same six shells in the same attachment conditions, assuming the part of the shell connected to the suction pad is fixed. In spite of this simplification, we find good agreement between the simulations and the experiments within $15\%$ maximal variation for the measured frequencies, with no adjustable parameters.

\begin{figure}[!h]
	\centering 		
	\includegraphics[width=0.8\textwidth]{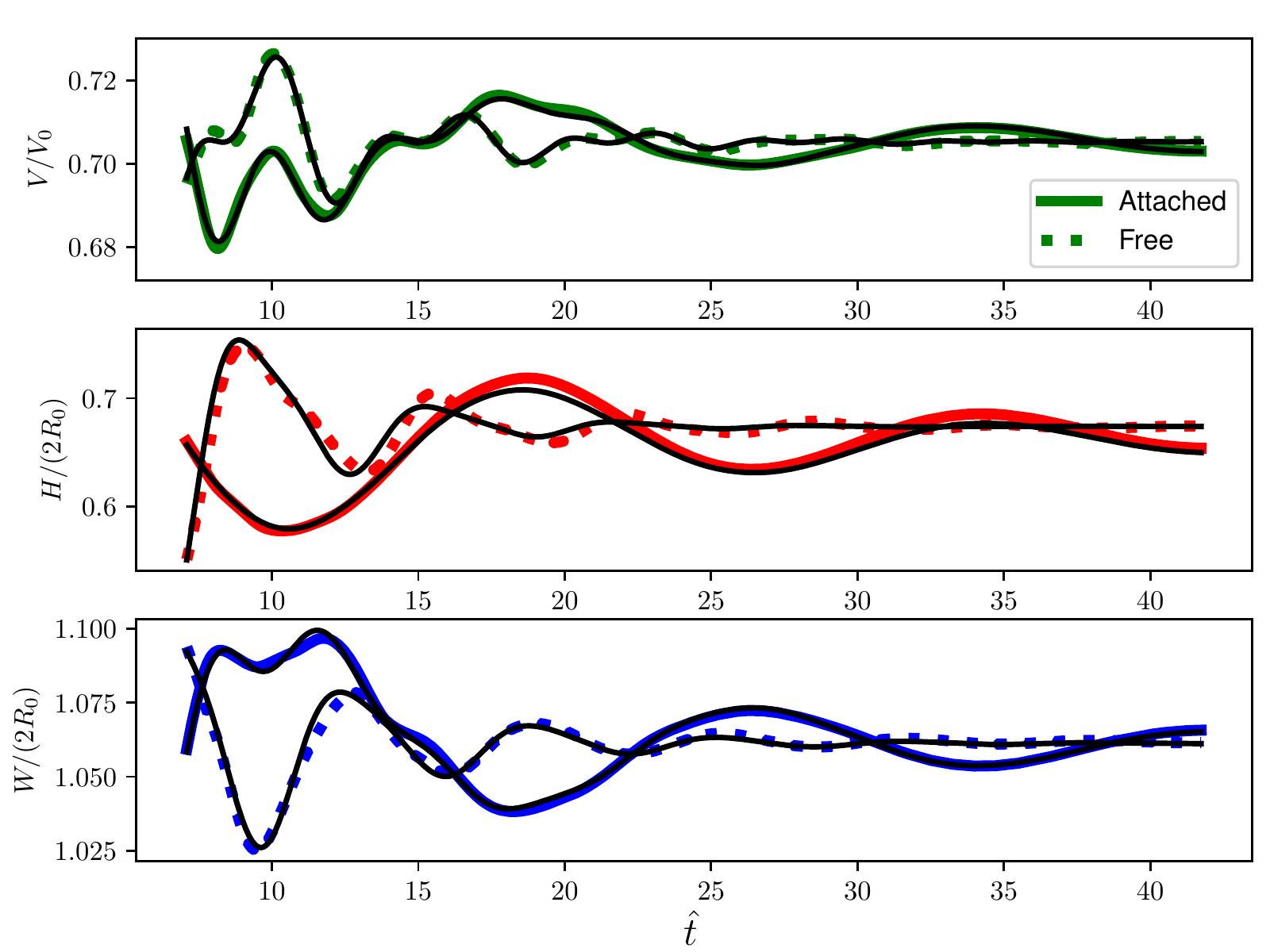}
	\caption[]{Oscillations of the volume $V$, height $H$ and width $W$ for a shell in configuration $\mathcal{R}$, which is attached or free (simulations). Thin black lines indicate the fit by the damped sum of two sinusoidal functions with the same frequencies $\omega_+$ and $\omega_-$ for the three space variables. The  amplitude ratio $A_+/A_-$ of the two functions differs according to the variable considered. Free case: $\hat{\omega}_+=2.0 $  and $\hat{\omega}_-=1.0 $ ; for $V$, $A_+/A_-=0.52$, for $H$, $A_+/A_-=0.21$, for $W$,  $A_+/A_-=0.22$.  Attached case: $\hat{\omega}_+=1.6 $  and $\hat{\omega}_-=0.4 $ ; for $V$, $A_+/A_-=1.3$, for $H$, $A_+/A_-=0.006$, for $W$,  $A_+/A_-=0.4$.\label{fig:detail-attached}}
\end{figure}

In addition, these simulations show that the oscillation frequency is smaller by 25 to 50 \%  than that of the free shell case. This is in qualitative agreement with the decrease of oscillation frequency of an initially free beam when it is attached at one end \cite{landau86}. Yet, this fact is not that intuitive since the attachment may lead to less fluid motion in the vicinity of the attachment patch, therefore less accelerated mass.  Noteworthy, the additional local constraint also modifies the relative weights of the two sinusoidal responses, depending on the considered quantity (Fig. \ref{fig:detail-attached}). While both frequencies are well present whatever the boundary condition when considering the volume oscillations, the picture is different when considering the oscillation of the height $H$ of the shell or of its width $W$, as defined in Fig. \ref{fig:exptyp}. In particular, for the height, a single frequency becomes predominant in the attached case, as seen in  Fig. \ref{fig:exptyp} for the experiments. As a signature of robustness of the modeling, the same pair of frequencies is sufficient to describe the oscillations whatever the considered quantity. A more systematic study of the dependency of each frequency on the boundary condition, and of its relative weight is the full signal, would require to define and parametrize the attachment area precisely. We simply conclude here that the two-oscillator modeling  is a robust description that remains valid for shells with fixed parts.

\section{Conclusion}

In spite of the potentially huge number of deformation modes of a buckled shell, its dynamics can be described by a pair of coupled oscillators.
 The two resulting modes decouple in the high inner pressure limit and can be identified as the surface oscillation mode and the volume oscillation mode. The former corresponds to the lowest frequency and contributes more to the overall signal than the high frequency mode. Both frequencies are much lower than that in the spherical case. This confirms and quantifies  the apparent softening of buckled shells observed in the literature. By contrast with previous models \cite{marmottant05,sijl11}, we show that this softening is mainly due to the absence of contribution of the inner gas. 
 
 We have also shown that over a large range of parameters, the dissipation mechanism is similar to that of a sphere in translation in a fluid, whose drag coefficient is Reynolds-dependent.
 
The established model can be used to anticipate the behavior of shells under more complex actuations than a fixed applied pressure. Using the model of Marmottant \cite{marmottant05}, it has recently been shown that buckling, by introducing an additional non-linearity, may trigger complex dynamics characterized by subharmonic behavior or even chaos \cite{Sojahrood11}. More work is also needed to identify what controls the relative weight of each mode, and how this weight will be influenced by the boundary conditions, like local bonds to another component. We plan to dig in that direction in a near future. The model could also be further enriched by incorporating the contribution of the bending energy, which may help  improving the collapse of data of Fig. \ref{fig:w} since it would introduce another dependency with the shell thickness. This contribution is already included in the description of equilibrium through the expression of the plateau pressure (Eq. \ref{eq:equiplateau}) but it is still poorly understood. Another open question is that of the viscous dissipation within the shell material.

\appendix

\section{Experimental set-up}
 \label{sec:setup}

As in Ref.\cite{djellouli17}, the spherical shells were realized by molding two half hollow spheres made of commercial elastomer resin "Dragon Skin$^{\copyright}$ 30" (Smooth-on) of Young modulus $E=0.5$ MPa (measured by traction experiments at 5\% elongation, as well as the Poisson's ratio $\nu = 0.5$). Other experiments on this material have shown that it behaves linearly at least until 25\% deformation rate \cite{Stein-montalvo21}, thus validating the use of a linear model in the simulation. The two halves were then glued together using the same material. Flat disks of diameter of order 20\% the shell diameter and thickness around 4\% the shell thickness were added at the pole of one of the two external moulds so as to create a weak point on one hemisphere, where buckling will systematically occur.

In order to explore the effect of size and of thickness over radius ratio, we have considered 6 different shells whose external radii $R_{\text{ext},0}$ lie between 7.5 and 75 mm while their dimensionless thicknesses $d/(R_{\text{ext},0}-d/2)$ are between  0.08 and 0.3. We cover one order of magnitude in size and the relative thickness was varied by a factor almost 4. Reaching broader ranges poses technical issues in the manufacturing process.

The shell under study was attached at the level of its pole opposite to the weak one to a fixed support  and immersed in a $60\times60\times60$ cm tank in anodized aluminum  with  polycarbonate polymer windows that could bear pressure differences of + 2 bar. The tank was filled with
 glycerol. Pressure variations were obtained with a pressure controller (OB1 by Elveflow)  connected to the thin layer of air that was left in the tank on top of the liquid. Shell deformations were recorded using a fast camera.  Pictures of an immersed grid of known characteristics allows to characterize the deformation of pictures due to window deformation by the overpressure. As in Fig. \ref{fig:exptyp}, the obtained images of the convex envelops of the shells are well contrasted and their contours  could be directly extracted and analyzed using home-made routines written in Python.

\section{Numerical method } \label{sec:method}

\begin{figure}[!ht]
	\centering 		
	\includegraphics[width=0.4\textwidth]{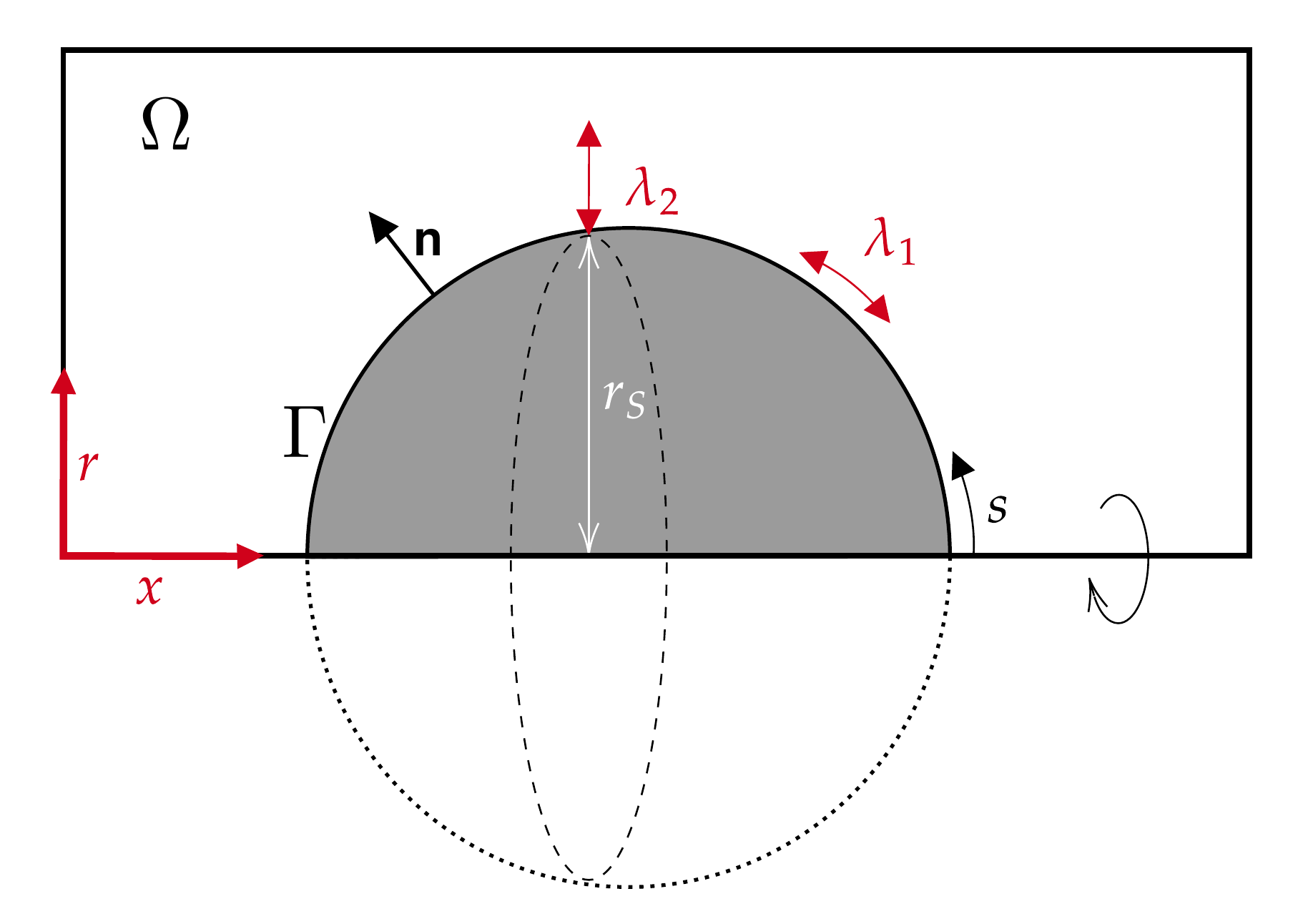}
	\caption[]{Numerical domain $\Omega$ for the simulations (white). The shell (dark gray) itself is not included. $\Gamma$ is the domain boundary referring to the shell surface. Illustration for the principal stretches in red.\label{fig:axisymmetric}}
\end{figure}
The problem is numerically solved with finite elements making usage of the finite element toolbox AMDiS \cite{Vey2007}. An axisymmetric arbitrary Lagrangian-Eulerian (ALE) method is employed according to \cite{mokbel2020}, where the membrane points move with the fluid velocity $\mathbf{v}$. This movement is harmonically extended to the internal part of the grid using one of the mesh smoothing approaches presented in \cite{mokbel2020}. 
%\subsubsection{Shell model}

The shell is modeled as an infinitely thin axisymmetric elastic surface, immersed in a Newtonian viscous fluid of viscosity $\eta_f$ and density $\rho_f$ and filled with air. The hydrodynamics in the surrounding fluid $\Omega$ are governed by the Navier-Stokes equations. The pressure due to the internal air is implemented as boundary conditions on the shell surface $\Gamma$. Assuming axisymmetric flow conditions reduces the problem to a 2D meridian $x-r-$ domain, describing half of the 3D domain's cross section. The simulation domain is shown in Fig.~\ref{fig:axisymmetric}. The problem in axisymmetric formulation reads \cite{abels2018,kim2005}
\begin{align}
\rho_f\left(\partial_t\mathbf{v} +\mathbf{v}\cdot\nabla\mathbf{v}\right) &=  \nabla\cdot \mathbf{S}  + \mathbf{v}_{\text{axi}} , & \text{in}\ \Omega \label{eq:NavierStokes}\\
\nabla\cdot\mathbf{v} + \frac{1}{r}v_r &= 0, & \text{in}\ \Omega \label{eq:incompressibility}\\
{\bf S} &= \eta_f (\nabla{\bf v}+\nabla{\bf v}^T) - P_f {\bf I}  &\text{in~}\Omega  
\end{align}
where ${\bf v}=(v_x,v_r)^T$ and $P_f$ are the fluid velocity and fluid pressure, respectively. The axisymmetric terms read
\begin{align}
\mathbf{v}_{\text{axi}} &= \left(\frac{2\eta_f}{r^2}v_r + \frac{1}{r}\left(\mathbf{S}+P_f\mathbf{I}\right)\right)\cdot  \begin{pmatrix} 0 \\ 1 \end{pmatrix}.\label{eq:axisymmetric}
\end{align}
The elastic shell reacts to in-plane (stretching) and out-of-plane (bending) deformations, quantified by the stretching energy $E_{\text{stretch}}$ and the bending energy $E_{\text{bend}}$. The stretching energy in an axisymmetric setting can be described in terms of the two principal stretches $\lambda_1$ and $\lambda_2$, which provide information about relative changes of surface lengths in lateral and circumferential direction, respectively. It is
\begin{align}
\lambda_1=\frac{\text{d}s}{\text{d}s_0}, \quad \lambda_2=\frac{r_{\rm S}}{r_{\text{S},0}},
\end{align}
with the arc length parameter $s$ of the interface curve $\Gamma$ and the distance of the shell $r_{\text{S}}$ to the symmetry axis. The subscript $_0$ refers to the quantities in initial state. The shell energies then read
\begin{align}
    E_{\text{bend}} &= \int_{\Gamma} 2K_B\left(H-H_0\right)^2\, \text{d}A\, \label{Eq:bend},\\
  E_{\text{stretch}} &= \int_\Gamma \frac{K_A+K_S}{2}\left(\left(\lambda_1-1\right)^2+\left(\lambda_2-1\right)^2\right) \nonumber\\
    & \qquad + \left(K_A-K_S\right)\left(\lambda_1-1\right)\left(\lambda_2-1\right) ~\text{d}A\, , \label{Eq:stretch}
\end{align}
with area bulk modulus $K_A$, area shear modulus $K_S$, bending stiffness $K_B$, mean curvature $H$ and spontaneous curvature $H_0$, which is the mean curvature in the initial state. These three quantities can all be calculated using Young's modulus $E$, the Poisson ratio $\nu$ (which will be set to $0.5$), and the (initial) thickness $d$ of the shell: 
\begin{align}
K_A = \frac{dE}{2(1-\nu)},\quad
K_S = \frac{dE}{2(1+\nu)},\quad
K_B = \frac{d^3E}{12(1-\nu^2)}\, .
\label{eq:moduli}
\end{align}
The force exerted by the shell is composed of the sum of the first variations of these energies and of the force due to the pressure of the internal air.  Note that our thin shell model is a simplification that is made for technical purpose, but the elastic model describes a shell of any thickness. Since the strains are small compared to the deformations of a thin shell, the in-plane force is linearized in $\lambda_i$ (see \cite{mokbel2017}, Appendix). In-plane and out-of-plane forces read \cite{mokbel2020}
\begin{align}
\frac{\delta E_{\mathrm{bend}}}{\delta\Gamma}  &= 4K_B\left[\Delta_{\Gamma}(H - H_0) + (4H^2 - 2K_g)(H-H_0) \right.\nonumber \\ & \quad\left. - 2H(H-H_0)^2\right]\textbf{n}\, ,\label{eq:bendingForce} \\
\frac{\delta E_{\mathrm{stretch}}}{\delta\Gamma} &=\left(2H\,\textbf{n}-\nabla_{\Gamma}\right)\left[\left(K_A+K_S\right)\left(\lambda_1 - 1\right) + \left(K_A-K_S\right)\left(\lambda_2 - 1\right)\right]\nonumber\\ &\quad + \frac{2K_S}{R}\left(\lambda_2-\lambda_1\right)\begin{pmatrix} 0 \\ 1 \end{pmatrix}\, , \label{eq:stretchingForce}
\end{align}
where $K_g$ is the Gaussian curvature. 

For the contribution of internal pressure, we assume homogeneous air pressure $P$ inside and use adiabatic gas theory to relate this pressure to the inner shell volume $V$ by $P = P_0\,(V_0/V)^{\kappa}$, where $P_0$ and $V_0$ denote the respective initial values, and $\kappa=1.4$ by assumption of an adiabatic process.
Accordingly, the stress exerted by the air is reduced to $P{\bf n} = P_0\,(V_0/V)^{\kappa}{\bf n}$ whereupon the boundary conditions become
\begin{align}
{\bf S} \mathbf{n} &= - P_0\left(\frac{V_0}{V}\right)^{\kappa}\mathbf{n}
- \frac{\partial E_{\text{bend}}}{\partial\Gamma}- \frac{\partial E_{\text{stretch}}}{\partial\Gamma}  & \text{on~}\Gamma \label{eq:BoundaryConditionsMicroswimmer}\\
P_f &= P_{\mathrm{ext}}& \text{on~}\partial\Omega\backslash\Gamma \label{eq:PextBoundaryCondition}%\\
%p_{\rm diff} &= p_{\rm ext}(t) - p_0\left(\frac{V_0}{V}\right)^{1.4}  & \text{on~}\Gamma  \\
%P_f &= p_{\text{prescribed}},& \text{on}~\partial\Omega\backslash\Gamma 
\end{align}

%\subsubsection{Discretization}
The numerical scheme in time step $m+1$ can be summarized as follows. An implicit Euler method is employed for the Navier-Stokes equations. In the following, time steps indices are denoted by superscripts. The scheme reads:
\begin{enumerate}
\item Calculate $\lambda_1, \lambda_2, H, K_g$ and ${\bf n}$ from the position of boundary grid points along $\Gamma$ (see \cite{mokbel2020} for details).
This allows computation of 
$\left(\frac{\delta E_{\mathrm{stretch}}}{\delta\Gamma}\right)^{m+1}$ and 
$\left(\frac{\delta E_{\mathrm{bend}}}{\delta\Gamma}\right)^{m+1}$ according to Eq.~\ref{eq:bendingForce} and Eq.~\ref{eq:stretchingForce}
\item Solve the Navier-Stokes equations Eq.~\ref{eq:NavierStokes}-Eq.~\ref{eq:axisymmetric} of the surrounding fluid. 
\item Move every grid point of $\Gamma$ with velocity ${\bf v}$.
\item Calculate the grid velocity ${\bf v_{\mathrm{grid}}}$ by using one of the mesh smoothing algorithms described in \cite{mokbel2020} to harmonically extend the interface movement into $\Omega$ 
and move all internal grid points of $\Omega$ with velocity ${\bf v_{\mathrm{grid}}}$.
\end{enumerate}

A function $P_{\rm ext}(t)$ imposes variable pressure at the boundary $\partial\Omega\backslash\Gamma$ of the computational domain, excluding the shell (see Eq.~\ref{eq:PextBoundaryCondition}). The values for $P_{\rm ext}$ start for $t=0$ at $P_0$. Afterwards, $P_{\rm ext}$ is increased slowly until the maximum desired value $P_{\rm max}$ is reached. The slow increase of pressure circumvents the occurrence of shell oscillations in the spherical compression stage before buckling. In order to analyze the post buckling oscillations, the maximum pressure value is kept constant afterwards. 

For the simulation results shown in Fig.~\ref{fig:resul}, we first compute the equilibrium state after buckling as explained above. Then, we impose an offset $\delta P$ to both, the internal pressure $P$ and the external pressure $P_{\text{ext}}$. The updated internal pressure acts as the new "initial" pressure $P_0$ for the rest of the simulation. Subsequently, a perturbation to the shell is applied by adding another prescribed pressure $P_{\text{perturb}}$ to the new $P_{\text{ext}}$. If the offset $\text{d}P$ is zero, we could confirm that these oscillations, whose initial amplitude is smaller, are similar to those obtained right after the shell has buckled, which initially implies larger deformation.

A weak spot is imposed into the formula when calculating the elastic moduli $K_A$, $K_S$, and $K_B$ by decreasing the thickness $d$ locally around the center of the weak spot, which is the membrane point touching the symmetry axis ($r=0$) on the right. This weak spot provides the perturbation needed to trigger buckling, otherwise our code is stable enough to let the shell stay in its spherical configuration, though unstable.

\begin{figure}
	\centering 		
\includegraphics[width=0.8\textwidth]{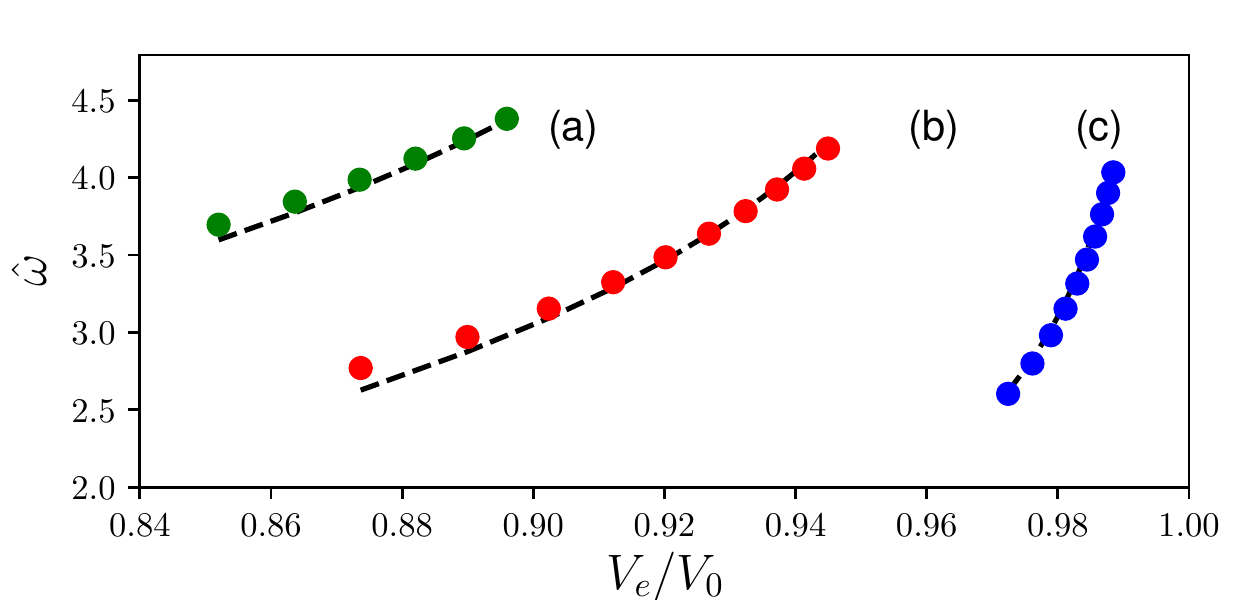}
	\caption[]{Pulsation $\hat{\omega}$ for spherical oscillation around equilibrium volume $V_e$, for the configuration $\mathcal{R}$ with varying initial pressures $\hat{P}_0$ that are equal inside and outside the shell and varying external pressure increases $\delta \hat{P}$. (a) and green: $\delta \hat{P}=0.45$ and $0.9<\hat{P}_0<1.8$ ; (b) and red: $\delta \hat{P}=0.225$ and $0.18<\hat{P}_0<1.8$ ; (c) and blue:  $\delta \hat{P}=0.045$ and $0.18<\hat{P}_0<1.8$. The dashed lines indicate the best fit with Eq. \ref{eq:wsphere}, where the pressure, geometric and elastic parameters are set to their values in the simulation and fixed, and the sole free parameter is a prefactor that will account for the mean error made in the simulation.\label{eq:oscillSph}}
\end{figure}

The oscillation frequencies are the main quantitative output of the simulations. They directly depend on the total accelerated mass and because of that, the size of the simulation box may bias the results. To estimate this, we consider the spherical oscillations of the shell under low pressure difference, such that it has not buckled.

To that purpose, we considered simulations of the shell in configuration $\mathcal{R}$ but considered different initial pressure inside and outside the shell $\hat{P}_0$  going from 0.18 to 1.8. Then we applied suddenly an external pressure $\hat{P}_{\text{ext}}=\hat{P}_0+\delta \hat{P}$ to the shell. For not too high values of $\delta \hat{P}$, the shell shrinks isotropically with damped oscillations towards its new configuration, characterized by an equilibrium volume $\hat{V}_{e}$.

The radial oscillations of a coated shell have been documented by several authors, for 2D shells (thin shells) \cite{deJong92,marmottant05} and for real shell of finite thickness \cite{church95,hoff00,doinikov06,chabouh21}. For 2D shells, the expressions for the oscillation frequency that are proposed in the literature are generally obtained considering oscillations around the stress-free configuration of volume $V_0$. Here, we are interested in oscillations around any state, which results in a modified expression for the pulsation, which we derive hereafter. In addition, models for thin shells generally consider only the stretching energy, which scales with $d$, and discard the curvature energy, which scales with $d^3$. Here, in Eq. \ref{eq:NRJsph}, we included both contributions in order to account for the oscillations of the mid-plane of shells of any thickness.

This elastic energy in the spherical configuration given in Eq. \ref{eq:NRJsph} can be rewritten in terms of $V$ and $V_0$, and we can follow the Lagrangian formalism that has lead to Eq. \ref{eq:eqmovV}. The pressure term is identical and the acceleration term can be calculated exactly by integrating the kinetic energy arising from the monopolar term from the equilibrium radius $R_e$ to $\infty$.

The pulsation in the absence of damping then reads

\begin{align}
    \omega_{\text{sph}}^2&=\frac{1}{\rho_f R_e^2}\,\Big[3 \kappa P_e+ \frac{4}{3} V_e^{-2/3}V_0^{-2/3}\,\Big(4 \pi K_b\,(2 V_0^{1/3}-V_e^{1/3})  \nonumber \\
    & \qquad +(36 \pi)^{1/3} K_a\, V_e^{1/3}(2 V_e^{2/3}-V_0^{2/3})\Big)\Big]. \label{eq:wsphere}
\end{align}

In the absence of bending energy, and for oscillations around the stress-free state ($V_e=V_0$), we recover the usual expression \cite{deJong92,marmottant05} $ \rho_f R_0^2 \omega_{\text{sph}}^2=3 \kappa P_0+4 K_a/R_0$.

In Fig. \ref{eq:oscillSph}, we plot the obtained pulsations in the simulation and compare them with the expected values obtained from Eq. \ref{eq:wsphere}. It is shown that the dependency with the shell parameters is recovered but the theoretical expression for $ \omega_{\text{sph}}$ has to be multiplied by a prefactor 1.07 to obtain an agreement between numerical data and theory.

This means that while the physics is well described by the simulations, they lead to an overestimation of the pulsation by $7\%$. This overestimation factor does not depend on the chosen configuration so we associate it with the finite size of the simulation box $\Omega$, which we checked by decreasing its size. The considered size in the paper is the result of a compromise between accuracy of the simulations and computation time.

\enlargethispage{20pt}

\dataccess{All data available from the authors upon reasonable request.}

\aucontribute{MM participated in the design of the study, developed the simulation code, carried out the numerical simulations, and critically revised the manuscript; AD performed the experiments and the image analysis  and critically revised the manuscript; CQ participated in the design of the study and critically revised the manuscript; SA developed the numerical model, participated in the design of the study and critically revised the manuscript. GC designed and coordinated the study, performed the data analysis, developed the theoretical model and drafted the manuscript. All authors gave final approval for publication and agree to be held accountable for the work performed therein.}

\competing{We declare we have no competing interests.}

\funding{SA acknowledges support from the German Research Foundation (grant AL 1705/3-2).}

\ack{Simulations were performed at the Center for Information Services and High Performance Computing (ZIH) at TU Dresden.
We thank G. Chabouh for introducing us with Refs. \cite{church95,renaud15}.
}

%%%%%%%%%% Insert bibliography here %%%%%%%%%%%%%%

\vskip2pc

%\bibliographystyle{RS} 

%\bibliography{biblioballons}

\end{document}